\documentclass[longauth]{aa}  
\usepackage{graphicx}
\usepackage{txfonts}
\usepackage{hyperref}
\usepackage{caption}
\usepackage{subcaption}
\usepackage{comment}
\usepackage{amsmath}
\usepackage{lscape}
\usepackage[dvipsnames,hyperref]{xcolor}

\hypersetup{
    colorlinks=true,
    linkcolor=Cerulean,
    citecolor=NavyBlue,
    filecolor=magenta,      
    urlcolor=cyan,
    pdftitle={Type-2 AGN},
    pdfpagemode=FullScreen,
    }

\newcommand{\NII}{{[N\,{\sc ii}]}}
\newcommand{\NIIs}{{[N\,{\sc ii}]\,}}
\newcommand{\NVs}{{[N\,{\sc V}]\,}}
\newcommand{\NVl}{{N\,{\sc V}\,$\lambda$}}

\newcommand{\NeIII}{{[Ne\,{\sc iii}]}}

\newcommand{\NeIV}{{[Ne\,{\sc iv}]}}

\newcommand{\NeIVl}{{[Ne\,{\sc iv}]\,$\lambda$}}
\newcommand{\NeV}{{[Ne\,{\sc v}]}}

\newcommand{\NeVl}{{[Ne\,{\sc v}]\,$\lambda$}}
\newcommand{\SII}{{[S\,{\sc ii}]}}
\newcommand{\SIIs}{{[S\,{\sc ii}]\,}}

\newcommand{\SIIll}{{[S\,{\sc ii}]\,$\lambda\lambda$}}

\newcommand{\OIII}{{[O\,{\sc iii}]}}
\newcommand{\OIIIs}{{[O\,{\sc iii}]\,}}
\newcommand{\OIIIl}{{[O\,{\sc iii}]\,$\lambda$}}
\newcommand{\OII}{{[O\,{\sc ii}]}}
\newcommand{\OIIs}{{[O\,{\sc ii}]\,}}

\newcommand{\OIIll}{{[O\,{\sc ii}]\,$\lambda\lambda$}}

\newcommand{\CIII}{{C\,{\sc iii}]}}
\newcommand{\CIIIs}{{C\,{\sc iii}]\,}}
\newcommand{\CIIIll}{\CIIIs$\lambda\lambda$}
\newcommand{\CIV}{{C\,{\sc iv}}}
\newcommand{\CIVs}{{C\,{\sc iv}\,}}
\newcommand{\CIVll}{{C\,{\sc iv}\,$\lambda\lambda$}\xspace}

\newcommand{\HeII}{{He\,{\sc ii}\,}}
\newcommand{\HeIIl}{{He\,{\sc ii}\,$\lambda$}}

\newcommand{\Ha}{H$\alpha$}
\newcommand{\Has}{H$\alpha$\,}
\newcommand{\Hb}{H$\beta$}
\newcommand{\Hbs}{H$\beta$\,}

\begin{document}

   \title{
   JADES: A large population of obscured, narrow line AGN at high redshift
    }
   \subtitle{}
   

\author{
Jan Scholtz
\inst{1,2}\fnmsep\thanks{Corresponding authors: js2685@cam.ac.uk \\}
\and
Roberto Maiolino
\inst{1,2,3}
\and
Francesco D'Eugenio
\inst{1,2}
Emma Curtis-Lake
\inst{4}
\and
Stefano Carniani
\inst{5}
\and
Stephane Charlot
\inst{6}
\and
Mirko Curti
\inst{7}
\and
Maddie S. Silcock
\inst{4}
\and
Santiago Arribas
\inst{8}
\and
William Baker
\inst{1,2}
\and
Rachana Bhatawdekar
\inst{9}
\and
Kristan Boyett
\inst{10,11}
\and
Andrew J. Bunker
\inst{12}
\and
Jacopo Chevallard
\inst{12}
\and
Chiara Circosta
\inst{9}
\and
Daniel J. Eisenstein
\inst{13}
\and
Kevin Hainline
\inst{14}
\and
Ryan Hausen
\inst{15}
\and
Xihan Ji
\inst{1,2}
\and
Zhiyuan Ji
\inst{14}
\and
Benjamin D. Johnson
\inst{13}
\and
Nimisha Kumari
\inst{16}
\and
Tobias J. Looser
\inst{1,2}
\and
Jianwei Lyu
\inst{14}
\and
Michael V. Maseda
\inst{17}
\and
Eleonora Parlanti
\inst{5}
\and
Michele Perna
\inst{9}
\and
Marcia Rieke
\inst{14}
\and
Brant Robertson
\inst{18}
\and
Bruno Rodr\'iguez Del Pino
\inst{9}
\and
Fengwu Sun
\inst{14}
\and
Sandro Tacchella
\inst{1,2}
\and
Hannah \"Ubler
\inst{1,2}
\and
Giacomo Venturi
\inst{5}
\and
Christina C. Williams
\inst{19}
\and
Christopher N. A. Willmer
\inst{14}
\and
Chris Willott
\inst{20}
\and
Joris Witstok
\inst{1,2}
}

\institute{
Kavli Institute for Cosmology, University of Cambridge, Madingley Road, Cambridge, 
CB3 0HA, UK\\
\and
Cavendish Laboratory, University of Cambridge, 19 JJ Thomson Avenue, Cambridge CB3 0HE, UK\\
\and
Department of Physics and Astronomy, University College London, Gower Street, London WC1E 6BT, UK\\
\and
Centre for Astrophysics Research, Department of Physics, Astronomy and Mathematics, University of Hertfordshire, Hatfield AL10 9AB, UK\\
\and
Scuola Normale Superiore, Piazza dei Cavalieri 7, I-56126 Pisa, Italy\\
\and
Sorbonne Universit\'e, CNRS, UMR 7095, Institut d'Astrophysique de Paris, 98 bis bd Arago, 75014 Paris, France\\
\and
European Southern Observatory, Karl-Schwarzschild-Strasse 2, 85748 Garching, Germany\\
\and
Centro de Astrobiolog\'ia (CAB), CSIC–INTA, Cra. de Ajalvir Km.~4, 28850- Torrej\'on de Ardoz, Madrid, Spain\\
\and
European Space Agency (ESA), European Space Astronomy Centre (ESAC), Camino Bajo del Castillo s/n, 28692 Villanueva de la Cañada, Madrid, Spain
\and
School of Physics, University of Melbourne, Parkville 3010, VIC, Australia\\
\and
ARC Centre of Excellence for All Sky Astrophysics in 3 Dimensions (ATRO 3D), Australia\\
\and
Department of Physics, University of Oxford, Denys Wilkinson Building, Keble Road, Oxford OX1 3RH, UK\\
\and
Harvard University, Center for Astrophysics $|$ Harvard \& Smithsonian, 60 Garden St., Cambridge 2138, USA\\
\and
Steward Observatory, University of Arizona, 933 N. Cherry Avenue, Tucson, AZ 85721, USA\\
\and
Department of Physics and Astronomy, The Johns Hopkins University, 3400 N. Charles St., Baltimore, MD 21218\\
\and
AURA for European Space Agency, Space Telescope Science Institute, 3700 San Martin Drive. Baltimore, MD, 21210\\
\and
Department of Astronomy, University of Wisconsin-Madison, 475 N. Charter St., Madison, WI 53706 USA\\
\and
Department of Astronomy and Astrophysics, University of California, Santa Cruz, 1156 High Street, Santa Cruz, CA 95064, USA\\
\and
NSF’s National Optical-Infrared Astronomy Research Laboratory, 950 North Cherry Avenue, Tucson, AZ 85719, USA\\
\and
NRC Herzberg, 5071 West Saanich Rd, Victoria, BC V9E 2E7, Canada\\
}

   \authorrunning{Scholtz, J et al.}
   \date{}

 
  \abstract
   {We present the identification of 41 narrow-line active galactic nuclei (type-2 AGN) candidates in the two deepest observations of the JADES spectroscopic survey with JWST/NIRSpec. The spectral coverage and the depth of our observations allow us to select narrow-line AGNs based on both rest-frame optical and UV emission lines up to z=10. Due to the metallicity decrease of galaxies, at $z>3$ the standard optical diagnostic diagrams (N2-BPT or S2-VO87) become unable to distinguish many AGN from other sources of photoionisation. Therefore, we also use high ionisation lines, such as \HeIIl4686, \HeIIl1640, \NeIVl2422, \NeVl3420, and \NVl1240, also in combination with other UV transitions, to trace the presence of AGN. Out of a parent sample of 209 galaxies, we identify 42 type-2 AGN (although 10 of them are tentative), giving a fraction of galaxies in JADES hosting type-2 AGN of about $20\pm5$\%, which does not evolve significantly in the redshift range between 2 and 10. The selected type-2 AGN have estimated bolometric luminosities of $10^{41.3-44.9}$ erg s$^{-1}$ and host-galaxy stellar masses of $10^{7.2-9.3}$ M$_{\odot}$. The star formation rates of the selected AGN host galaxies are consistent with those of the star-forming main sequence. The AGN host galaxies at z=4-6 contribute $\sim$18-30 \% to the UV luminosity function across different UV luminosity bins, slightly increasing with UV luminosity. }


   \maketitle
%
\section{Introduction} \label{sec:intro}

It has been widely accepted that supermassive black holes (SMBHs) reside in the centre of most (perhaps all) massive galaxies. During their accretion phases, SMBHs are observed as active galactic nuclei \citep[AGN;][]{Rees82, LyndenBell69, Soltan82,Merloni04}. The tight correlation between a SMBH mass and host-galaxy bulge properties (such as velocity dispersion and mass) at $z\sim0$ indicates a strong connection between the growth of a SMBH and its host galaxy
\citep[e.g.,][]{Magorrian98, Kormendy13}. Furthermore, (radiative and mechanical) feedback from AGNs is a key ingredient of galaxy evolution, as AGNs can inject a significant amount of energy into the interstellar and circumgalactic medium (ISM and CGM, respectively) of their host galaxies. AGN feedback is necessary to reproduce key galaxy properties such as: colour bi-modality, galaxy sizes, and a broader range of specific star formation rates and enrichment of the intergalactic medium (IGM) by metals, as well as the high-mass drop-off of the stellar mass function compared to the halo mass function \citep[e.g.,][]{Silk98,DiMatteo05,Alexander12,Dubois13,Dubois13b,Vogelsberger14,Hirschmann14, Crain15,Segers16,Beckmann17,Harrison17,Choi18,Scholtz18}.

The identification and study of AGN at high redshift is essential to understanding not only the co-evolution of SMBHs and galaxies, but also the formation of galaxies at early epochs. Although over the past 15 years there has been significant progress in the identification of active SMBHs at high redshift ($z>3$)  \citep[e.g.][]{Merloni10, Bongiorno14, Trakhtenbrot17,Mezcua2018, Lyu2022}, the majority of these detections have been limited to bright quasars identified in large-volume ground-based surveys (e.g., \citealt{Banados16,Shen19}, see \citealt{Inayoshi20, Fan22} for a review). These include the most distant identified quasars at $z\sim7.5$ \citep{Banados18,Yang20,Yang21}. 

With the launch of the \textit{James Webb Space Telescope} \citep[JWST;][]{Gardner2023,Rigby2023}, we now have access to rest-frame UV-to-optical emission lines of galaxies up to $z\sim12$. These lines allow us to identify and study AGNs, even at low masses and luminosities, via optical and UV diagnostic diagrams. A number of AGN candidates have already been identified by performing Spectral Energy Distribution (SED) analyses of broad-band photometry from NIRCam and MIRI aboard JWST \citep[][]{Furtak22,Onoue23,Barro23,Yang23,Bogdan23,Ignas23, Lyu23, Yang2023}. Significant progress has been made using deep spectroscopy from JWST/NIRSpec \citep{Boker22, Jakobsen22} and the JWST/NIRCam grism, tracing the presence of a broad line region \citep[BLR; ][]{furtak23,greene23, Harikane23, Kocevski23, Maiolino23gnz11,Maiolino23JADES, Matthee23, Onoue23,Ubler23}. These observations revealed a previously unseen population of AGN at z$>4$, and out to z$\sim$11, with estimated black hole masses (M$_{\rm BH}$) in the range $\rm 10^6$ to $\rm 10^8 ~M_\odot$ and bolometric luminosities $\rm 10^{44}-10^{45} ~erg~s^{-1}$. These black hole masses and luminosities are 2--3 orders of magnitude lower than those inferred for quasars at the same redshifts \citep{Mazzucchelli23, Zappacosta23}.

However, the identification of narrow-line (i.e. type-2, as opposed to type-1 AGN showing BLR emission) AGN has remained unexplored with JWST data. This has been the main AGN identification tool at low redshift, however, so far has had limited success at high redshift. The reason is most likely the rapid evolution of the metallicity and ionisation parameter at z>3 towards more metal-poor and higher ionisation parameter \citep[][]{Hirschmann22, Curti23JADES, Curti23ERO,Tacchella23, Trump23}. Indeed, \citet{Harikane23, Kocevski23, Maiolino23JADES, Ubler23} have shown that the type-1 AGN found by JWST have narrow line ratios on the classical emission-line diagnostic diagrams (such as BPT; \citealt{BPT1981,Kauffmann03,Kewley13} and VO87; \citealt{VO87}) that overlap with the local SF sequence, and not with the region occupied by nearby and low-z AGN. This displacement has been predicted by photoionization models for low metallicity AGN \citep{Groves06, Nakajima22}. Additionally, \citet{Feltre16} and \citet{Gutkin16} ran photo-ionisation grid models to assess the effect of low metallicity and high ionisation parameter on rest-frame optical and UV emission lines, and found that the nebular emission of star-forming galaxies at high redshift become similar to that of AGN, further complicating the AGN selection via narrow emission line diagnostics. The identification of type-2 AGN at high redshift can still rely on high ionization lines, such as the \NeIVl2424 (with ionisation potential $>$63.45eV) detected by \citet{Brinchmann23, Chisholm24} in a galaxy at z = 7.66. Luckily, the identification of type-2 AGN with high metallicity through standard BPTs is still possible (see \citealt{Perna23b})

Despite these newly discovered AGN being far less luminous than the previously known quasar population and being hosted in galaxies with lower masses, they can play a major role in shaping galaxy evolution \citep[via positive and negative feedback; ][]{Koudmani21,Koudmani22} and could potentially significantly contribute to the reionisation of the Universe. Indeed, we have already observed these processes in GN-z11 \citep{Bunker23gnz11} believed to host an AGN \citep{Maiolino23gnz11} with detected p-cygni CIV] outflow in the system, while \citet{Scholtz23} observed heating and ionisation of the circum-galactic medium around this unique galaxy. These examples show the presence of AGN feedback at high z, which can explain the early emergence of quiescent galaxies and high burstiness of star formation \citep{Carnall23a,Carnall23b,Endsley23,Looser23a,Looser23b,Dome23,Strait2023}; indeed, \citet{Gelli23} showed that supernovae feedback is not sufficient to rapidly suppress the star formation in some of these rapidly quenched systems.

In this paper, we leverage two of the deepest spectroscopic observations from JWST Advanced Deep Extragalactic Survey (JADES, Proposal ID: 1210 \& 3215; \citealt{Bunker23JADES, Curtis-lake23,Eisenstein23,Eisenstein23_3215, Robertson23}). Using this deep spectroscopy with JWST/NIRSpec, we aim to identify type-2 AGN using deep spectroscopy with JWST/NIRSpec in galaxies with stellar masses M$_{*}\sim 10^{6.5}-10^{9.5}$ M$_{\odot}$ from redshift z = 1 and out to the highest redshifts for which rest-frame UV and optical nebular lines are accessible to NIRSpec (z$\sim$12.0). We reassess the demarcation lines between AGN and star-forming galaxies at high redshift and further refine selection criteria using the photo-ionisation modelling introduced by \citet{Feltre16} and \cite{Nakajima22}. 

The paper is organized as follows. In Section \ref{sec:obs}, we describe the observations, data reduction, spectral fitting procedures, and SED analysis to derive stellar masses and star formation rates (SFRs). In Section \ref{sec:selection} we identify AGN using their narrow line properties of rest-frame UV and optical emission lines. In Section \ref{sec:discussion}, we estimate the properties of the AGN and their host galaxies and discuss our results. In Section \ref{sec:conclusions}, we draw our conclusions. Throughout this paper, we use the AB magnitude system and assume a flat $\Lambda$CDM cosmology with $\Omega$m = 0.315 and H$_{0}$ = 67.4 km/s/Mpc  \citet{Planck2020} cosmology and a \citet{Chabrier03} initial mass function.

\section{Observations, Data reduction and Analysis}\label{sec:obs}

The observations of our targets were obtained as part of the JADES survey, utilising the multi-object spectroscopic capabilities of the \emph{JWST}/NIRSpec micro-shutter array (MSA; \citealt{Jakobsen22, Ferruit22}) across two programmes: PID 1210 \& PID 3215.

Five disperser/filter combinations were used: the low-resolution PRISM/CLEAR \citep[$0.6<\lambda<5.3~\mu$m$, R=30-300$;][]{Jakobsen22}, the medium-resolution gratings G140M/F070LP, G235M/F170LP and G395M/F290LP ($0.6<\lambda<5.3~\mu$m, $R=1,000$), and the high-resolution grating G395H/F290LP ($2.8<\lambda<5.1~\mu$m, R=2,700). For the 1210 program, the observations consist of individual visits with a per-visit duration of 33.6~ks for the prism, and of 8.4~ks for each grating. Each galaxy was assigned a minimum of one and a maximum of three visits depending on its priority \citep{Bunker23JADES}. This resulted in a maximum integration time of $\sim100$~ks for the prism and of 25~ks for each grating. The 3215 program consisted only of three configurations: PRISM/CLEAR, G140M/F070LP, and G395M/F290LP, with a maximum resulting integration time of 168~ks, 42~ks, and 168~ks.


The observations were performed in the three-shutter nod mode, so that common targets were observed in different shutters and different locations on the detector. Therefore, each visit required a unique MSA configuration. Each target allocation (performed using the
eMPT tool; \citealt{Bonaventura23}\footnote{\url{https://github.com/esdc-esac-esa-int/eMPT_v1}}) was designed to maximise the number of targets that have all the disperses/filters in common between all the dither positions, however, not all targets have the full on source integration time outlined above. In total, we observed 481 unique targets.

\begin{figure*}
    \centering
    \includegraphics[width=0.9\paperwidth]{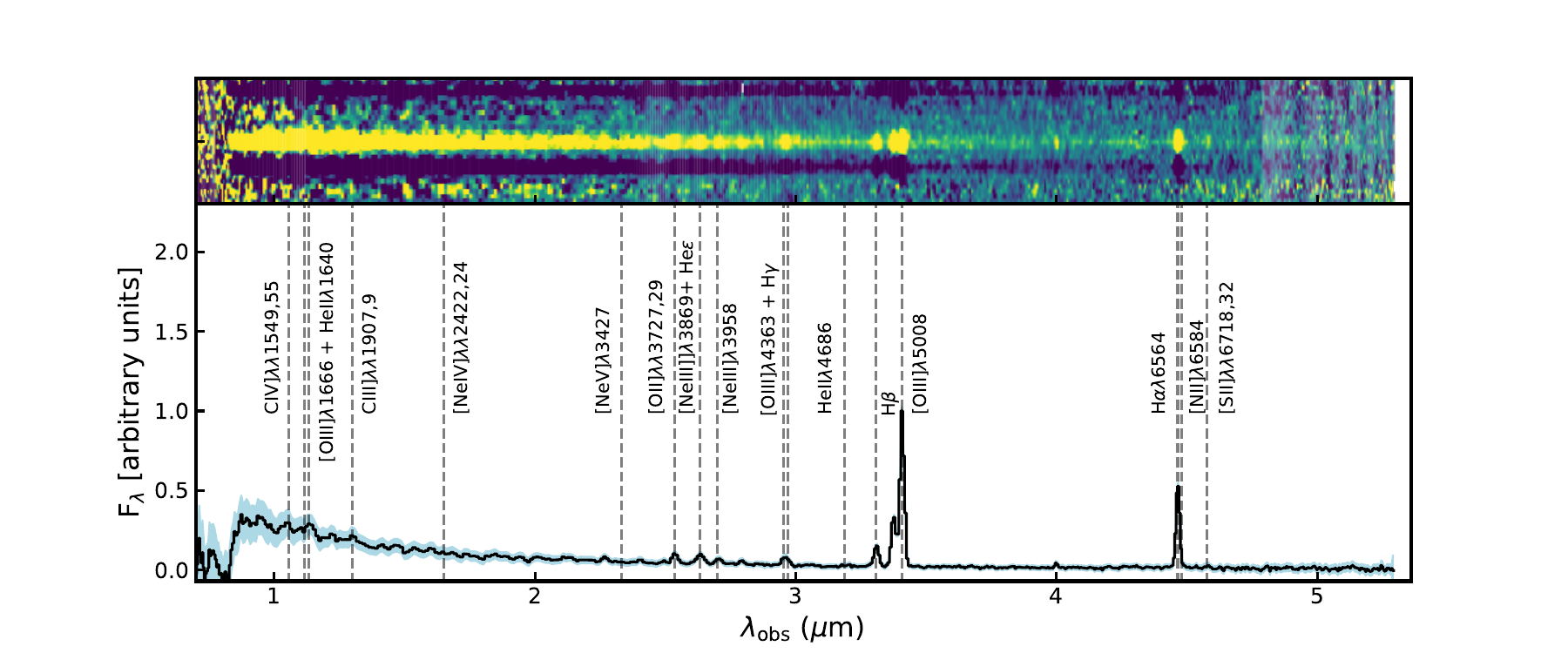}
   \caption{Example of low-resolution PRISM/CLEAR spectrum of galaxy (JADES-NS-GS-00022251) at z = 5.804 included in our parent sample. The spectrum is shown in units of F$_\lambda$ normalized to the \OIIIl5008 peak. We highlight major emission lines used in our analysis. 
    }
   \label{fig:Examples}
\end{figure*}

The MSA configurations have been designed to avoid spectral overlap for the prism mode. However, since the spectra taken with the medium or high-resolution gratings occupy a significantly larger portion of the detector, and to avoid removing targets with overlapping spectra, we allowed spectral overlap for these modes. The MSA configurations were, however, designed to minimize the negative effect of spectral overlap on our science, since the highest priority targets are not allowed to be contaminated by neighboring spectra. 
We show examples of the acquired PRISM/CLEAR spectra in Figure \ref{fig:Examples}.

\subsection{Data Reduction}\label{sec:data_red}

The JWST/NIRSpec MSA observations were processed with the data reduction pipeline of the ESA NIRSpec Science Operations Team (SOT) and the NIRSpec GTO Team, further detailed description of the GTO pipeline will be presented in Carniani et al. (in prep), and is summarised in \citet{Bunker23JADES}. Here, we briefly summarize the procedure. We retrieve the level-1a data products from the MAST archive and estimate the count-rate slopes per pixel, using the unsaturated groups in the ramps. We remove any jumps in the ramps due to cosmic rays by estimating the slope of each ramp. During this stage, we perform the master dark and bias subtraction, as well as the flagging of saturated pixels. The background subtraction is performed pixel by pixel by combining the three nod exposures of each pointing. We note that for some targets we excluded one of the 3-shutter nods in the background subtraction stage as a  serendipitous source contaminated the open shutters. We performed the flat-field correction of the spectrograph optics and disperser corrections on the 2D-dimensional cutouts of each of the three-shutter slits. The path loss corrections are calculated assuming point sources and taking into account the source location on the shutter. 

Finally, the individual 2D maps are interpolated onto an irregular wavelength grid for the PRISM/CLEAR observations (to avoid oversampling the line spread function below 2 $\mu$m) and onto a regular grid for the gratings. We extracted the 1D spectra from the 2D maps adopting a box-car aperture centered on the relative position of the targets. We combined all 1D spectra and removed the bad pixels by adopting a sigma-clipping approach.

\subsection{Spectral fitting using PPXF}\label{sec:fitting}

As the continuum is well detected in all our targets in the PRISM observations, we employed \texttt{pPXF} \citep{Cappellari2017, Cappellari2022} to fit the continuum and emission lines simultaneously. As we are interested in inactive galaxies or type-2 AGN, there is no contamination of the continuum by the AGN, making \texttt{pPXF} an ideal tool for this analysis. The continuum is fitted as a linear superposition of simple stellar-population (SSP) spectra, using non-negative weights and matching the spectral resolution of the observed spectrum.
As input stellar templates we used the synthetic library of simple stellar population
spectra (SSP) from \textsc{fsps} \citep{conroy+2009,conroy_gunn_2010}. This library uses
MIST isochrones \citep{choi+2016} and C3K model atmospheres \citep{conroy+2019}. We
also used fit a 5\textsuperscript{th}-order multiplicative polynomial, to capture the 
combined effects of dust reddening, residual flux calibration issues, and any systematic
mismatch between the data and the input stellar templates.
We find that the NIRSpec/MSA data are fit adequately by low-order polynomials. Increasing the polynomial degree does not improve the fit results (as quantified by the value of the reduced $\chi^2$).
To simplify the fitting, any flux with a wavelength shorter than Lyman break is manually set to 0. The full description of the \texttt{pPXF} fitting will be presented in D'Eugenio et al. (in prep). 

For the emission lines fitting we use the redshifts published in\citet{Bunker23JADES}, which used redshifts determined from the medium resolution gratings available and PRISM spectra otherwise. All emission lines are modeled as single Gaussian functions, matching the observed spectral resolution. We use vacuum wavelengths for the emission lines throughout the paper. In order to remove degeneracies in the fitting and reduce the number of free parameters, the emission lines are split into four separate kinematic groups, bound to the same redshift and \emph{intrinsic} broadening. These groups are as follows:
\begin{itemize}
    \item UV lines with rest-frame $\lambda< 3000$~\AA. 
    \item The hydrogen Balmer series.
    \item Non-hydrogen optical lines with rest-frame $\lambda< 9000$~\AA. 
    \item Near infra-red lines. 
\end{itemize}
The stellar kinematics are tied to the Balmer line kinematics. For emission line multiplets arising from the same level, we fixed the emission line ratio to the value prescribed by atomic physics (e.g., \OIIIl5008/\OIIIl4959 = 2.99). For multiplets arising from different levels, the emission line ratio can vary. This is relevant for five multiplets: \CIVll1548,1551, \CIIIll1907,1909, \OIIll3727,3730 and \SIIll6718,6733. However, in practice, the spectral resolution of the PRISM/CLEAR observations is insufficient to resolve the individual lines in each multiplet. For this reason, we model each of these as a single Gaussian in the PRISM observations. In addition, as the \HeIIl1640 and \OIIIl$\lambda$1661,66 are blended together we are unable to use their measured fluxes from the PRISM observations and instead, choose to use the R1000 observations.
The fits are performed in two steps. In the first run, we tie the kinematics as described above, and identify robust detections (above 5-$\sigma$ significance) for re-fitting. In the second run, we only fit for these detected lines, but allow for their kinematics to vary independently from one another.

The R1000 gratings were fitted using the same procedure as for the PRISM observations, by combining all three gratings and fitting them simultaneously. However, we did not stack the spectra from different gratings, because this would combine the highest-resolution end of one grating with the poorest-resolution start of the next grating.

\subsubsection{Fitting the UV lines}\label{sec:UV_fitting}

The high ionisation UV lines were excluded from the \texttt{pPXF} fitting as they are extremely faint and blended in the PRISM data. Since the AGN high ionisation UV lines are considerably fainter than the optical emission lines, we fitted only objects that are well-detected in [OIII] or H$\alpha$ emission (SNR$>10$). 
 
We fitted the \NeIVl2424, \NeVl3420 and \NVl1240 using  \texttt{QubeSpec}'s\footnote{\url{https://github.com/honzascholtz/Qubespec}} fitting module. Each emission line was fitted using a single Gaussian component and the continuum was fitted as a power law. This simplistic approach is sufficient for describing a narrow range of the continuum around an emission line of interest. As \texttt{QubeSpec} is a Bayesian code implemented using \texttt{emcee} \citep{emcee}, it is necessary to also supply priors on each quantity. The peak of the Gaussian and continuum normalization are given a log-uniform prior, while the FWHMs are set to a uniform distribution spanning from the minimum resolution of the NIRSpec/MSA ($\sim$200 kms$^{-1}$) up to a maximum of 1200 km/s. The prior on the redshift was a normal distribution centered on the redshift obtained from \texttt{pPXF} with a standard deviation of 300 km/s. This is done because as the high ionisation UV lines originate from close to the accretion disc, there can be a significant velocity offset between the low and high ionisation lines in an AGN. We report the fluxes of each of these high ionisation lines in the Appendix in Table \ref{table:Sample_high_ion}. Throughout this work we derived upper limits as 3$\sigma$. 

\subsection{SED fitting using BEAGLE}\label{sec:SED_fitting}

In order to compare our selected AGN with the rest of the galaxy population, it is necessary to measure the stellar masses and star-formation rates of our sample. We use the full fitting of the slit-loss corrected PRISM spectra using the \texttt{BEAGLE} tool \citep{Chevallard2016}. We assume a delayed-exponential star-formation history (SFH), while decoupling the current SFR from the previous SFH by allowing a recent duration of 10Myr of constant star formation to vary independently. A \citet{Chabrier03} initial mass function (IMF) with an upper mass cut-off of 100 M$_{\odot}$ was adopted using the updated \citet{bruzual03} stellar population models described in  \citet{Vidal-Garcia17}. We define the total stellar mass as the mass currently locked into stars. This definition accounts for the fraction of mass returned to the ISM during stellar evolution. The SFRs of our objects are averaged over 10 million years.

We note that the SFRs of the whole sample estimated with \texttt{BEAGLE} are in excellent agreement with those estimated from the dust-corrected Balmer lines (H$\alpha$ and H$\beta$; see \citealt{Curti23JADES} for more information). 

\subsection{Comparison to photoioinization models}\label{sec:cloudy}

At high redshifts, galaxies become more metal-poor and show an increasingly higher ionisation parameter  \citep[see][]{Hirschmann22,Schaerer2022,Cameron23,Curti23JADES, Trump23}, resulting in standard diagnostics diagrams used to identify AGNs at low redshift becoming less useful \citep[see][]{Kewley13, kewley19, Hirschmann22}. We, therefore, consider the photoionisation models initially described in \citet{Feltre16} and \citet{Gutkin16}, and updated with more recent stellar spectra and with a better description of AGN cloud microturbulence \citep{VidalGarcia17,Mignoli19,Hirschmann19}. These works consider a large grid of photoionisation models computed using the \texttt{CLOUDY} code \citep{Ferland2013} for star formation and AGN narrow-line regions, and for various gas metallicities, dust content, and ISM densities. From these model grids, we selected all models with metallicities between 0.001 and 0.02 (corresponding to 0.06-1.3 solar) and a dust-to-metal mass ratio of 0.3. This value is intermediate between the range observed in the most metal-poor absorbers \citep[e.g.,][]{konstantopoulou+2023} and the Milky-Way value of 0.45. We consider all models with carbon-to-oxygen abundance ratio in the range 0.38-1.00 solar to describe a variety of different or less common star-formation grids. We further restrict the grids to only include models with an IMF upper mass cut-off of 300 M$_{\odot}$, similar to the SED fitting above. 

We also compare our observed emission line ratios with the models of \citet{Nakajima22}. They investigated the emission line ratios of star-forming galaxies, AGN, PopIII stars, and Direct Collapse black holes (DCBHs)  using the \texttt{Cloudy} code and the BPASS stellar population models \citep{Eldridge2017}. In this work, we only use the models for star-forming galaxies and AGN host galaxies, as our objects do not have low enough metallicities to host either PopIII or DCBHs \citep[based on values from ][]{Curti23JADES}. We selected the same metallicities, dust-to-metal mass ratio and IMF upper cutoff as for the \citet{Gutkin16} and \citet{Feltre16} models.

Throughout this work, we present these models on our diagnostics plots (Figures \ref{fig:BPT_full}, \ref{fig:HeII_full}, \ref{fig:UV_lines} and \ref{fig:NeIII_OII}) as light blue and yellow circles for AGN and star-formation, respectively. We will further discuss these points and use them to redefine the selection of AGN based on to-be-introduced N2-BPT and S2-VO87 diagnostic diagrams in \S \ref{sec:optical}.

\section{Selection of AGN in JADES deep spectroscopic data}\label{sec:selection}

\subsection{The parent sample }\label{sec:parent}

In order to select AGN based on their narrow line properties we first need to define the parent sample. In this work, we use data from JADES surveys - programmes 1210 (the JADES HST-Deep tier) and 3215 (the JADES Origins Field). As discussed in \cite{Bunker23JADES}, \cite{DEugenio2024_DR3} and \cite{Eisenstein2023JOF}, the sources for spectroscopic follow-up were selected in a somewhat heterogeneous way, also as a consequence of matching priorities with the densities of sources, for an optimal allocation of the shutters. However, generally, the selection of high-z candidates was mostly based on UV/optical rest-frame selection, magnitude-limited, and priority decreasing with redshift.

For obscured, type 2 AGN, there is no obvious reason that this parent sample is biased against or in favour of AGN, as the continuum is dominated by the host galaxy stellar light. One may expect that the strong emission lines associated with AGN may boost the flux in some bands, or preferentially favour the redshift identification. However, at high redshift `normal' star-forming galaxies are typically strong line emitters with high equivalent widths, hence AGN should not be preferentially selected or identified in these parent samples.

\begin{table}
 \centering
   \caption{Definitions of line ratios adopted throughout the paper.}
 \begin{tabular}{ll}
  \hline
  Diagnostics & Line Ratio\\
  \hline
  R3 & [\ion{O}{III}]$\lambda$5008 / H$\beta$\\
  N2 & [\ion{N}{II}]$\lambda$6584 / H$\alpha$\\
  S2 & [\ion{S}{II}]$\lambda\lambda$6718,32 / H$\alpha$\\
  Ne3O2 & [\ion{Ne}{III}]$\lambda$3869 / [\ion{O}{II}]$\lambda\lambda$3727,29\\
  He2 & \ion{He}{II}$\lambda$4686 / H$\beta$\\
  C43 &  \ion{C}{IV}$\lambda\lambda$1549,51/\ion{C}{III]}$\lambda\lambda$1906,08\\
  C3He2 & \CIII$\lambda\lambda$1906,08 / \ion{He}{II}$\lambda$1640\\
  Ne4C3 &  \NeIVl$\lambda$2422,24/\ion{C}{III]}$\lambda\lambda$1906,08\\
  Ne5C3 &  \NeVl$\lambda$3427,29/\ion{C}{III]}$\lambda\lambda$1906,08\\
  N5C3  &  \NVl$\lambda$1239,42/\ion{C}{III]}$\lambda\lambda$1906,08\\
 \hline
 \end{tabular}
  \label{table:line_ratios_def}
\end{table}

\begin{sidewaystable*}
 \centering
   \caption{List of AGN from JADES HST Deep selected in this work. Columns: ID, redshift, detection method $^{+}$, stellar mass, SFR and UV magnitude from BEAGLE, AGN bolometric luminosity and any other additional notes. We note that for bolometric luminosity the uncertainties are dominated by the systematical uncertainties (0.3-0.5 dex).}
    \begin{tabular}{@{}lcccccccc@{}} 
\hline 
\hline 
ID & field &z& selection & log$_{10}$ M$_{*}$ & SFR & M$_{\rm UV}$ & log$_{10}$ L$_{\rm bol}$  & Notes\\
  &  & & method    & M$_{\odot}$       &M$_{\odot}$ yr$^{-1}$ & & ergs s$^{-1}$ \\
 \\
\hline 
JADES-NS-GS00004902& 1210 &  5.123&S2-VO87*&  8.6$^{0.03}_{0.02}$&  1.4$^{0.21}_{-0.14}$& -19.0&  43.0&\\
JADES-NS-GS00007099& 1210 &  2.860&S2-VO87*&  9.1$^{0.01}_{0.01}$&  1.1$^{0.07}_{-0.06}$& -17.4&  42.9&\\
JADES-NS-GS00007762& 1210 &  4.146&High ion&  8.2$^{0.02}_{0.02}$&  4.1$^{0.29}_{-0.28}$& -18.8&  43.6&outflows\\
JADES-NS-GS00008083& 1210 &  4.665&High ion \& HeII$\lambda4686$&  7.8$^{0.01}_{0.01}$&  7.3$^{0.12}_{-0.12}$& -18.5&  44.5&Type-1 \& LAE\\
JADES-NS-GS00008456& 1210 &  1.884&HeII$\lambda4686$&  8.1$^{0.03}_{0.08}$&  0.1$^{0.03}_{-0.01}$& -16.1&  41.6&\\
JADES-NS-GS00008880& 1210 &  2.327&N2-BPT&  7.1$^{0.25}_{0.70}$&  0.1$^{0.03}_{-0.02}$& -14.9&  42.0&\\
JADES-NS-GS00009422& 1210 &  5.942&HeII$\lambda1640$ \& HeII$\lambda4686$&  7.7$^{0.00}_{0.00}$&  5.4$^{0.04}_{-0.03}$& -19.8&  44.5&LAE\\
JADES-NS-GS00009452& 1210 &  5.135&N2-BPT&  8.1$^{0.17}_{0.16}$&  4.2$^{1.17}_{-0.97}$& -17.9&  43.6&\\
JADES-NS-GS00010073& 1210 &  2.632&HeII$\lambda4686$&  6.9$^{0.06}_{0.04}$&  0.8$^{0.05}_{-0.04}$& -16.5&  43.0&\\
JADES-NS-GS00016745& 1210 &  5.574&S2-VO87*&  8.3$^{0.03}_{0.04}$&  6.5$^{0.26}_{-0.27}$& -19.5&  43.7&\\
JADES-NS-GS00017072& 1210 &  4.707&HeII$\lambda1640$&  8.2$^{0.06}_{0.15}$&  0.4$^{0.11}_{-0.06}$& -17.9&  42.3&\\
JADES-NS-GS00017670& 1210 &  2.350&HeII$\lambda4686$&  8.4$^{0.01}_{0.01}$&  0.9$^{0.04}_{-0.04}$& -17.8&  43.0&\\
JADES-NS-GS00021842& 1210 &  7.981&High ion&  7.4$^{0.03}_{0.02}$&  2.4$^{0.08}_{-0.07}$& -18.6&  43.8&LAE\\
JADES-NS-GS00022251& 1210 &  5.804&HeII$\lambda1640$&  7.9$^{0.03}_{0.03}$&  4.7$^{0.12}_{-0.13}$& -18.9&  44.1&\\
JADES-NS-GS10000626& 1210 &  4.468&HeII$\lambda4686$ \& High ion&  6.8$^{0.29}_{0.20}$&  0.3$^{0.01}_{-0.01}$& -16.8&  42.5&\\
JADES-NS-GS10008071& 1210 &  2.227&S2-VO87&  8.8$^{0.01}_{0.01}$&  7.2$^{0.15}_{-0.17}$& -17.9&  43.1&\\
JADES-NS-GS10011849& 1210 &  2.686&S2-VO87*&  8.1$^{0.03}_{0.04}$&  2.5$^{0.06}_{-0.06}$& -18.0&  43.6&\\
JADES-NS-GS10012477& 1210 &  0.665&S2-VO87&  7.8$^{0.01}_{0.02}$&  0.1$^{0.00}_{-0.00}$& -15.7&  41.5&\\
JADES-NS-GS10012511& 1210 &  2.019&N2-BPT&  8.2 $^{0.3}_{-0.4}$&  3.2 $^{0.7}_{-0.9}$& -16.5&  41.7&\\
JADES-NS-GS10013597& 1210 &  3.320&HeII$\lambda4686$&  7.5$^{0.13}_{0.06}$&  0.8$^{0.20}_{-0.12}$& -17.4&  42.6&\\
JADES-NS-GS10013609& 1210 &  6.931&High ion&  7.7$^{0.06}_{0.07}$&  3.9$^{0.19}_{-0.18}$& -18.7&  44.0&outflows\\
JADES-NS-GS10013905& 1210 &  7.206&HeII$\lambda4686$&  7.4$^{0.05}_{0.03}$&  2.2$^{0.14}_{-0.12}$& -18.6&  43.7&\\
JADES-NS-GS10015338& 1210 &  5.073&HeII$\lambda4686$&  7.9$^{0.04}_{0.04}$&  3.6$^{0.15}_{-0.19}$& -19.4&  43.9&LAE\\
JADES-NS-GS10035295& 1210 &  3.588&HeII$\lambda1640$&  7.4$^{0.07}_{0.03}$&  2.3$^{0.06}_{-0.05}$& -18.0&  43.7&\\
JADES-NS-GS10036017& 1210 &  2.016&N2-BPT \& S2-VO87&  9.3$^{0.01}_{0.02}$&  85.5$^{1.69}_{-1.95}$& -20.7&  44.1&\\
JADES-NS-GS10040620& 1210 &  1.776&N2-BPT&  8.1$^{0.04}_{0.04}$&  0.1$^{0.02}_{-0.01}$& -16.4&  41.7&\\
JADES-NS-GS10056849& 1210 &  5.821&High ion&  7.1$^{0.06}_{0.04}$&  1.1$^{0.05}_{-0.04}$& -18.1&  43.2&LAE\\
JADES-NS-GS10058975& 1210 &  9.437&High ion \& HeII$\lambda1640$&  8.1$^{0.03}_{0.03}$&  6.6$^{0.20}_{-0.20}$& -20.3&  44.4&\\
JADES-NS-GS00095256& 3215 &  4.159&S2-VO87&  7.4$^{0.12}_{0.33}$&  0.3$^{0.21}_{-0.14}$& -16.9&  43.0&\\
JADES-NS-GS00099671& 3215 &  5.936&HeII$\lambda4686$&  7.5$^{0.05}_{0.05}$&  1.2$^{0.07}_{-0.06}$& -18.0&  42.9&\\
JADES-NS-GS00104075& 3215 &  3.717&S2-VO87*&  7.2$^{0.12}_{0.07}$&  1.3$^{0.29}_{-0.28}$& -17.1&  43.6&\\
JADES-NS-GS00108487& 3215 &  3.975&S2-VO87&  8.9$^{0.02}_{0.03}$&  0.5$^{0.12}_{-0.12}$& -18.2&  44.5&\\
JADES-NS-GS00111091& 3215 &  4.497&S2-VO87&  8.6$^{0.03}_{0.04}$&  0.4$^{0.03}_{-0.01}$& -17.4&  41.6&\\
JADES-NS-GS00111511& 3215 &  3.008&S2-VO87*&  8.0$^{0.03}_{0.05}$&  0.4$^{0.03}_{-0.02}$& -16.8&  42.0&\\
JADES-NS-GS00114573& 3215 &  2.881&S2-VO87*&  8.0$^{0.04}_{0.05}$&  0.2$^{0.04}_{-0.03}$& -17.3&  44.5&\\
JADES-NS-GS00132213& 3215 &  3.012&S2-VO87&  7.7$^{0.04}_{0.03}$&  0.3$^{1.17}_{-0.97}$& -17.6&  43.6&\\
JADES-NS-GS00143403& 3215 &  0.735&S2-VO87*&  8.2$^{0.01}_{0.01}$&  0.1$^{0.05}_{-0.04}$& -15.4&  43.0&\\
JADES-NS-GS00201127& 3215 &  5.837&S2-VO87*&  8.2$^{0.06}_{0.08}$&  2.1$^{0.26}_{-0.27}$& -18.5&  43.7&\\
JADES-NS-GS00202208& 3215 &  5.450&HeII$\lambda1640$&  8.1$^{0.02}_{0.02}$&  11.7$^{0.11}_{-0.06}$& -19.7&  42.3&\\
JADES-NS-GS00208643& 3215 &  5.566&HeII$\lambda4686$&  7.2$^{0.01}_{0.01}$&  1.5$^{0.04}_{-0.04}$& -18.0&  43.0&\\
JADES-NS-GS00209979& 3215 &  1.883&S2-VO87*&  8.5$^{0.01}_{0.01}$&  1.8$^{0.08}_{-0.07}$& -17.9&  43.8&\\
\hline 
\end{tabular} 

    \par $^{+}$ high-ion refers to identification based on \NeIVl2424 and \NeVl3420. $^{*}$ Tentative selection
  \label{table:Sample}

\end{sidewaystable*}

\subsection{Selecting AGN based on optical emission lines}\label{sec:optical}

From these two samples with require a 3$\sigma$ detection of the following lines: \Ha, \Hb, \OIIIs and \CIII, and wavelength coverage of \SII, \NIIs and  UV and optical \HeII \space and \CIV\  for the optical and UV line selection. Overall, this yields a sample size of 110 and 99 sources for the programmes 1210 and 3215, respectively.

We define the line ratios used in this paper in Table \ref{table:line_ratios_def} (see below) and we summarise all emission lines, their wavelengths and ionisation potential in the Appendix in Table \ref{table:eml}.

We note that the emission line fluxes used in the diagnostics diagrams used below are not corrected for dust obscuration. However, this effect is minimal as we are selecting emission line ratios that are generally close to each other in the wavelength. Indeed for the average A$_{\rm V}$ = 0.34 of our sample (from Beagle SED fitting; see \S~\ref{sec:SED_fitting}), the emission line ratio changes by less than 0.15 dex. For any emission line diagnostics where this ratio changes by more than 0.1 dex (\NeVl/\CIII and \NVl/\CIII; Figure \ref{fig:High_ion}), we add an arrow to indicate the direction and magnitude of dust attenuation assuming the median value of the sample. We note that the dust attenuation does not change our selection of the AGN in our sample.

We consider a galaxy an AGN candidate if it is classified an AGN in at least one diagnostic. Furthermore, we report the list of selected AGN candidates, field, coordinates, redshift, the method with which we detect it, and other notes in Table \ref{table:Sample}.

\begin{figure*}
    \includegraphics[width=0.95\textwidth]{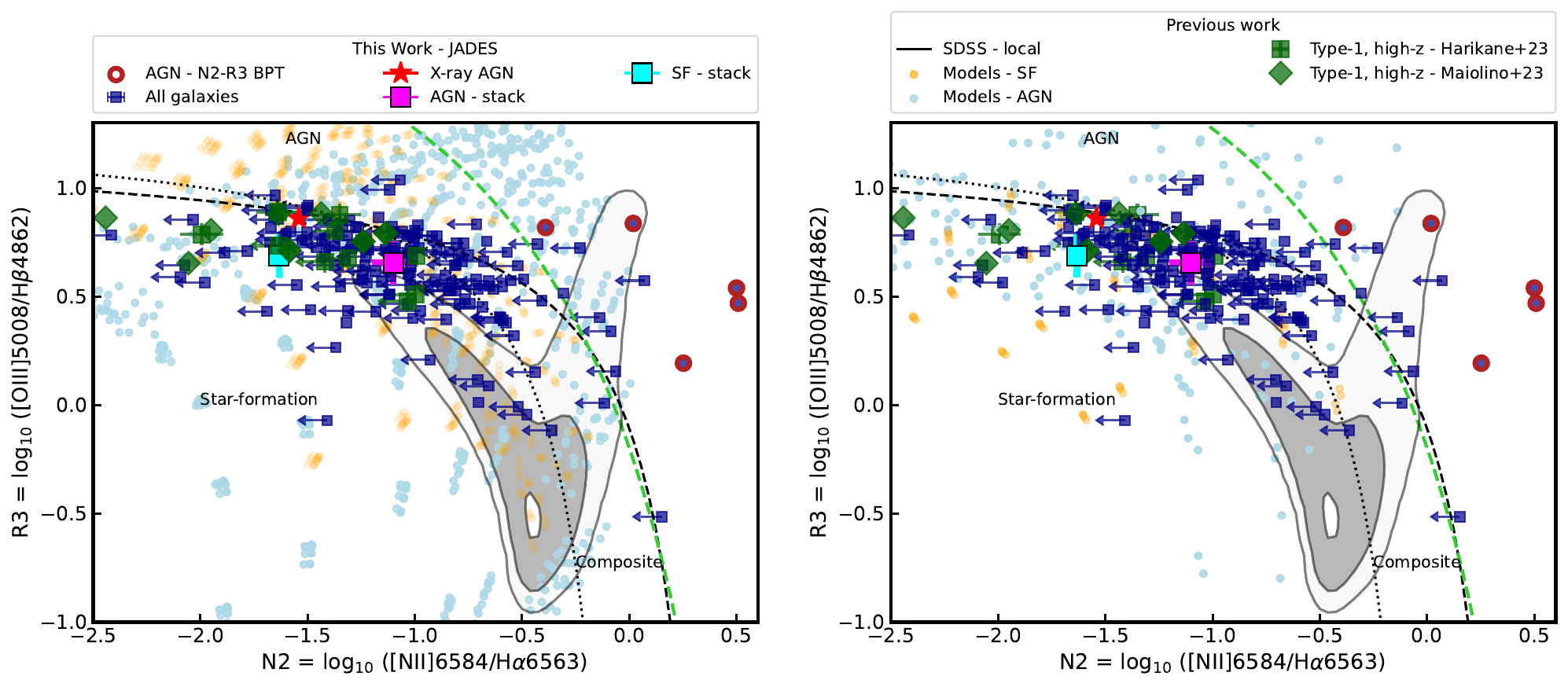}
    \includegraphics[width=0.95\textwidth]{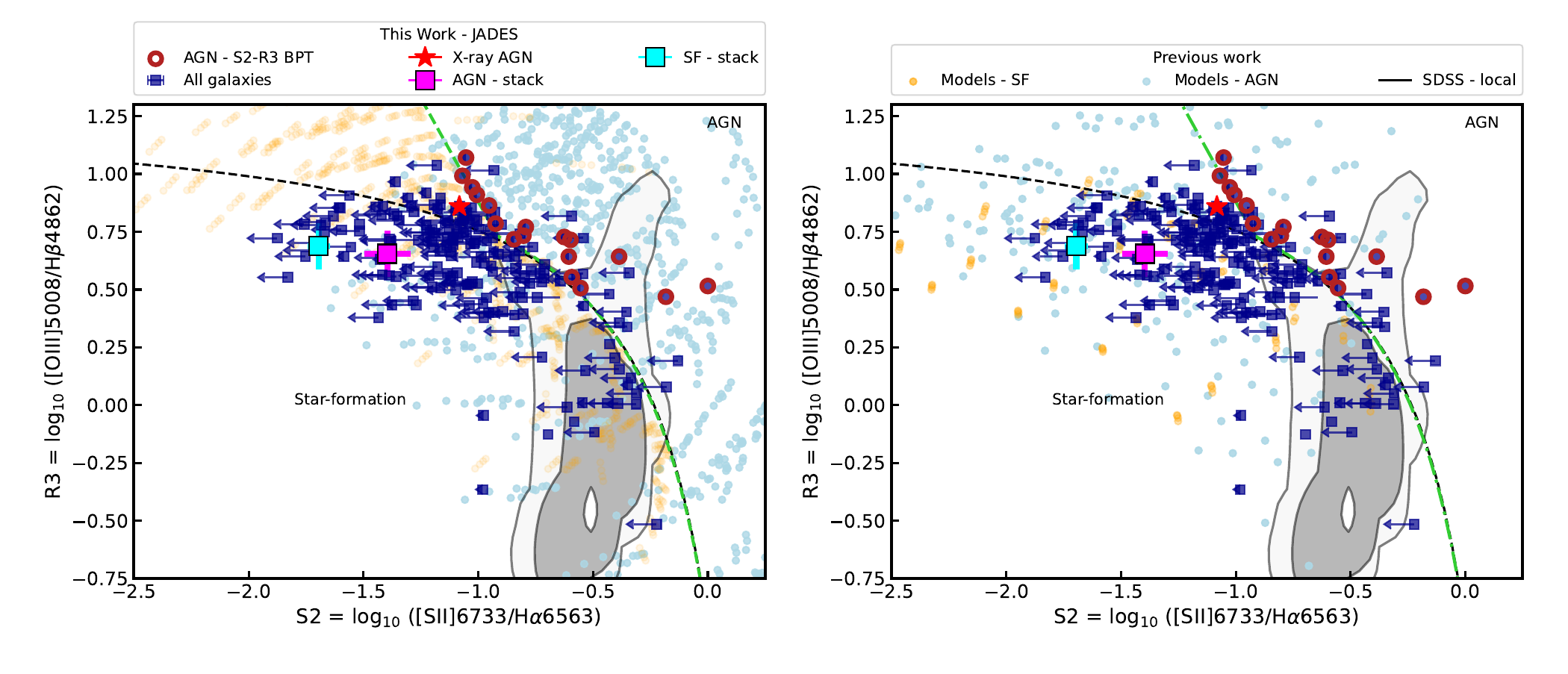}
   \caption{ Typical line ratio diagnostic diagrams used to select AGN. N2-R3 BPT (\NII/\Ha \space vs \OIII/\Hb; top row) and S2-VO87 (\SII/\Ha \space vs \OIII/\Hb; bottom row). The ionisation models of \citet{Feltre16,Gutkin16} (left) and \citet{Nakajima22} (right) are reported as yellow (SF) and blue (AGN) points (see \S \ref{sec:cloudy}). The objects from this work are plotted as blue squares. The objects selected as AGN based in these diagrams are highlighted by red circles. We also plot our new demarcation lines as green dashed. The black dashed lines show the star forming versus AGN demarcation lines from \citet{Kewley01} and \citet{Kauffmann03}. For comparison, we plot  SDSS galaxies shown as a grey contour plot. The magenta and cyan squares show a stacked spectrum for AGN and star-forming galaxies (see \S \ref{sec:stacking}). We highlight type-1 AGN from \citet{Harikane23} and \citet{Maiolino23JADES} as green squares and diamonds. The red star shows the X-ray selected AGN in our sample. 
     }
   \label{fig:BPT_full}
\end{figure*}

We plot our sources as dark blue points on the N2--R3 and S2--R3 planes (also known as BPT and VO87 diagrams, respectively; \citealt{Baldwin81,VO87}) in the top and bottom row of Figure \ref{fig:BPT_full}, respectively. The low metallicities of high-z objects lead to faint \NII\ emission lines, which are typically undetected at z$>$4 \citep{Cameron23, Curti23JADES}. Furthermore, the high ionisation parameter pushes star-forming sources to high R3 values, towards the classical demarcation between star formation and AGN defined by \citet{Kewley01, Kauffmann03, Kewley13}.
On the other hand, the low metallicity of the Narrow Line Region at high-z results in
the line ratios of AGN to shift away from the locus of AGN typically populated by
AGN and move towards the locus of star-forming galaxies
 \citet{Nakajima22,Harikane23, Maiolino23JADES}. The combination of these effects
 makes high-z AGN and SF galaxies to largely overlap on these diagrams and makes
 the selection of high-z AGN much more challenging than in the local Universe.
 However, these diagrams can still be used to identify AGN via a conservative
 selection approach, as discussed below

In Figure \ref{fig:BPT_full}, we plot results from different photoionisation models for a wide range of gas properties (metallicities, ionisation parameter and densities) in the left and right panels, with star-forming models and AGN models shown in yellow and blue points, respectively.  The left panels show AGN models from \citet{Feltre16} and star-forming models from \citet{Gutkin16}, while the right panels show the AGN and star-forming models of \citet{Nakajima22}. The photo-ionisation models indeed show that star-forming galaxies can lie in the AGN part of the BPT diagram at high redshift with log$_{10}$(\OIII/\Hb)$>0.9$ (as also noted in Figure 14 of \citealt{Feltre16}). As such, we need a clean selection to account for galaxies with low metallicity and high ionisation parameters. To identify a conservative demarcation line between star-forming galaxies and AGN, we define the edge of the star-forming region using the points with the highest \NII/\Has values from either the \citet{Kewley01} line or \citet{Gutkin16} models. We clarify that this is a very conservative method to select AGN. Galaxies above this demarcation line can be safely classified as AGN, however, as discussed above, certainly there are plenty of AGN also mixed with the galaxy population below this line, especially at these high redshifts. Indeed the type-1 AGN from \citet{Harikane23, Maiolino23JADES} do lie in the star-formation part of the BPT diagram. 

We fit these points, marking the edge of the star-forming region of the BPT diagram, with the functional form as in \citet{Kewley01, Kauffmann03}:
\begin{equation}
    Y= \frac{a}{(X-b)} + c
\end{equation}
where Y = log$_{10}$([OIII]/H$\beta$) and X=log$_{10}$([NII]/H$\alpha$). We report the results for the demarcation line in Table \ref{table:new_fits}. In the BPT diagram, this functional form well describes the edge of star-forming galaxies and we show this line as a green dashed line in the top panels of Figure \ref{fig:BPT_full}. We select five AGN based on the new demarcation line, all at z<5. We highlighted these sources with red circles in the top panel Figure \ref{fig:BPT_full}. 

We repeat the same analysis on the S2-R3  diagram (S2-VO87) to make new demarcation lines between AGN and star-forming galaxies. However, the previous functional form no longer fits the edges of the \citet{Kewley01} line and \citet{Gutkin16} points and we need to adapt it as:
\begin{equation} \label{eq1}
\begin{split}
Y = &  \frac{a}{(X-b)} + c \;[X>-0.92]\\
  = & d+eX   \;[X>-0.92]
\end{split}
\end{equation}
We report the parameters of the AGN demarcation lines in Table \ref{table:new_fits}. Overall, we detect seventeen AGN in the S2-R3 BPT diagram, using the \citet{Kewley01} or \citet{Gutkin16} line (dashed green line in the bottom panels of Figure \ref{fig:BPT_full}). We note that there are several AGN candidates close to the demarcation line. We discuss these selection methods and their reliability in \S \ref{sec:AGN_sel_discussion}.

\begin{table}
 \centering
   \caption{Fitted parameters for equation 1 when fitted for the N2-R3 and S2-R3 diagnostics plot. We define the edge of star-forming galaxies described in \S\ref{sec:optical}. }
 \begin{tabular}{lccccc}
  \hline
  Model & a & b & c & d & e \\
  \hline
  \hline
  N2-R3 & 3.08 &  1.03 &  2.78 & - & -\\ 
    \hline
  S2-R3  & 0.78 & 0.34  &  1.36 & -0.91 & -1.79  \\ 
    \hline 
 \end{tabular}
  \label{table:new_fits}
\end{table}

\begin{figure*}
    \includegraphics[width=0.9\paperwidth]{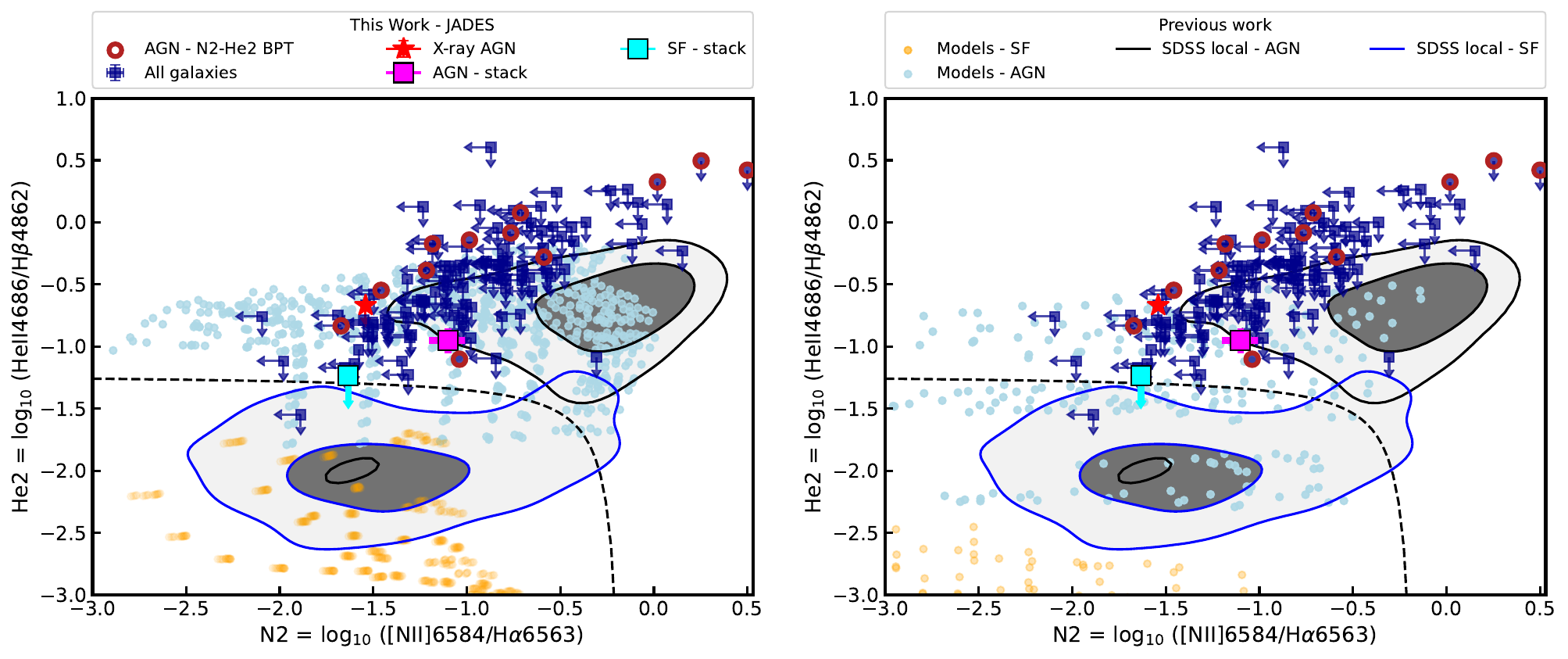}
   \caption{ The He2-N2 (\HeII/\Hb \space vs \NII/\Ha) diagram for the sample of JADES galaxies for our sample, plotted as blue squares. The left and right plots show ionisation models from \citet{Feltre16,Gutkin16} and \citet{Nakajima22}, respectively, as yellow and light blue points (see \S \ref{sec:cloudy}). The magenta and cyan squares show a stacked spectrum for AGN and star-forming galaxies (see \S \ref{sec:stacking}). The black dashed line indicates a demarcation line between star-forming and AGN galaxies by \citet{Shirazi12}. The blue and black contours show the star-forming galaxies and AGN from SDSS, respectively. We highlight the selected AGN in this diagram with a red circle. 
    }
   \label{fig:HeII_full}
\end{figure*}

A number of studies \citep{Shirazi12, Baer17, Dors23, Tozzi23} investigated the identification of AGN in SDSS using the \HeIIl4686 emission line. This is a recombination line whose flux is nearly independent of the gas metallicity and ionization parameter, depending instead primarily on the shape of the ionizing spectrum and, more specifically, on the number of ionizing photons with energies beyond 54~eV. We plot the He2-N2 diagram in Figure \ref{fig:HeII_full}. In this case, we do not redefine the demarcation line as defined by \citet{Shirazi12} as the original line is more conservative than the photo-ionisation models used in our work. Overall, we detect \HeIIl4686 in nine galaxies, whose HeII/H$\beta$ ratio indicates the presence of hard ionising radiation indicating AGN. Despite the deep JWST observations, the upper limits on \HeIIl4686/H$\beta$ do not provide any constraints on the presence of an AGN. We note that we selected three additional sources solely based on their N2 ratio $>0.2$ (see Figure \ref{fig:BPT_full}). Although these sources do not have \HeIIl4686 detections, their large N2 ratio is already identified with the N2-BPT. In summary, based on this diagram we select 12 AGN in total.

We finally note that the only X-ray detected AGN in our sample (red star in Figures \ref{fig:BPT_full} and \ref{fig:HeII_full}) is located in the star-forming region of the N2-BPT and S2-VO87 diagnostic diagrams \citep[hence it would have been missed by this classification, as most type 1 AGN, as discussed in ][]{Harikane23,Maiolino23JADES,Kocevski23}, while the upper limit on the HeII$\lambda$4686 would no identify as AGN, further showing complexity of selecting AGN at high redshift. 

\subsection{Selecting AGN based on UV line diagnostics}\label{sec:UV_lines}

\begin{figure*}
    \includegraphics[width=0.9\paperwidth]{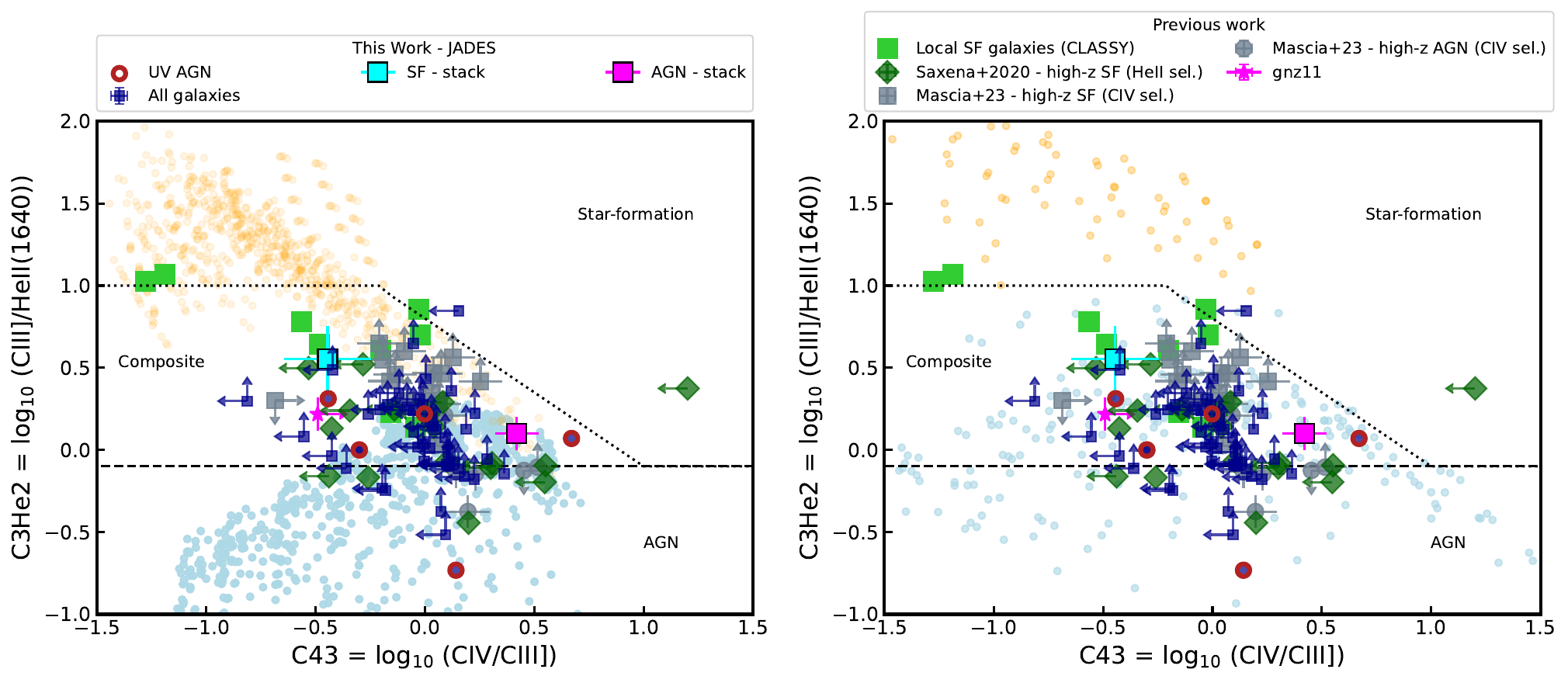}
   \caption{ \CIII/\HeIIl1640 vs \CIV/\CIII  (CHe2-C43) diagnostic diagram.  We plot our sample as blue squares. The left and right plots show ionisation models from \citet{Feltre16,Gutkin16} and \citet{Nakajima22}, respectively, as yellow and light blue points (see \S \ref{sec:cloudy}). The green diamonds show \HeIIl1640 detection from the VANDELS survey \citep{Saxena20}, and all \CIV\space detections from \citet{Mascia23} as grey colour with squares and circles representing SF and AGN, respectively. The green squares show local analogues of high-redshift galaxies \citep[from CLASSY survey;][]{Mingozzi23}. We also show the black dashed and dotted demarcation lines between AGN, star-forming galaxies and composite line ratios from \citet{Hirschmann22}. Overall, we select five AGN on this diagram and we highlight these with a red circle. The magenta and cyan squares show a stacked spectrum for AGN and star-forming galaxies (see \S \ref{sec:stacking}).
    }
   \label{fig:UV_lines}
\end{figure*}

In the previous section, we showed that the selection of AGN at high-z using the BPT and S2-VO87 diagrams is extremely challenging for all but metal-rich galaxies. In this section, we will focus on the UV emission lines. The clearest signature of AGN activity is high-ionisation lines such as NV$\lambda$1240, \NeIVl2424, \NeVl3426 \citep[][]{Feltre16}. They require high energy photons (77, 63 and 97 eV, respectively), that can hardly be produced by star-formation processes and are not seen even in galaxies hosting WR stars \citep{Mingozzi23}; hence, as already mentioned, these are clear signatures of AGN. However, they are often very faint even in powerful AGN \citep{Ubler23}. 

Past works on this topic have suggested various UV emission lines ratios based on \CIII$\lambda\lambda$1907,09, \CIV$\lambda\lambda$1548,51, \OIIIs1660,66 and \HeIIl1640 \citep[][]{Feltre16, Hirschmann22, Mingozzi22, Mascia23}. These lines are ideal in identifying AGN at z$>6.8$, where most rest-frame optical emission lines are redshifted out of the wavelength range of JWST/NIRSpec. However, even in luminous AGN the majority of these lines are beyond detectability for a  4-7 hour JWST/NIRSpec exposure at high redshift.

Therefore, in this work, we focus mainly on the brightest emission lines of the UV emission lines that are potentially detectable in our data, or on which we can at least obtain meaningful upper limits. In Figure \ref{fig:UV_lines}, we show the ratio \CIII$\lambda\lambda$(1906+1908)/\HeIIl1640 (C3He2) against \CIV1548,50/\CIII$\lambda\lambda$(1906+1908) (C43). We also plot the models of AGN and star-forming galaxies as blue and orange points, respectively, from \citet{Feltre16, Gutkin16} (left plot) and \citet{Nakajima22} (right plot). The black and dotted lines show the demarcation lines between star-forming galaxies, AGN, and composite objects derived by \citet{Hirschmann22}. They obtain such a line from the post-processing of the Illustris TNG cosmological simulation using the models of \citet{Gutkin16} and \citet{Feltre16}. In our sample, there are six targets for which we detect \HeIIl1640. Out of these five objects, one lies in the AGN part of the diagram. Furthermore, four targets are in the composite part of the diagram according to \citet{Hirschmann22}, but they are classified as  AGN based on \citet{Feltre16} and \citet{Nakajima22}.

We also show the objects from \citet{Saxena20} with \HeIIl1640 detection in the VANDELS survey \citep{VANDELS18} as green diamonds. This sample removed all X-ray-detected objects, removing any signs of obvious AGN. However, most of the type-1 AGN detected with JWST do not show any X-ray emission even in the deepest observations (Maiolino et al. in prep), indicating that many of these VANDELS sources might have a significant AGN contribution. All the VANDELS sources would be classified as AGN by \citet{Feltre16, Gutkin16, Nakajima22}, whereas 14 of them fall in the AGN or composite region defined by \citet{Hirschmann22}. We also show \CIV\space detections from VANDELs survey \citep{Mascia23} as grey squares and circles representing SF and AGN host (identified from X-ray or as type-1 AGN) galaxies, respectively. The AGN and SF galaxies from \citet{Mascia23} are well separated on our diagram as predicted by the photo-ionisation models used in our work. 

We compare our observations with local high-redshift analogues from the CLASSY survey \citep{Berg22, Bethan22}, specifically the data from \citet{Mingozzi23}. The CLASSY sources are classified as star-forming galaxies based on the BPT and \HeIIl4686 optical emission lines. Overall, the five objects in our sample with \HeIIl1640 selected as AGN in our sample have significantly lower C3-He2 ratios than other "pure" star-forming galaxy samples, confirming our AGN selection. 

We note that \citet{Mascia23} devised a classification of UV-BPT using the VANDELS survey. However, most of the diagnostics diagrams used in that work use both \CIV\space and \HeIIl1640 in the same emission line ratio. Unfortunately, the majority of our sample is undetected in both \CIV\space and \HeIIl1640 which does not allow us to place these objects on their diagnostics diagram. We will further investigate these diagnostics with the full JADES sample in future works. 

We note that the above emission line diagnostic requires the detection of three UV emission lines. Despite the excellent sensitivity of JWST/NIRSpec, detecting all three emission lines is still a challenge for exposures $<$10 hours. 

\begin{figure}
    \includegraphics[width=0.99\columnwidth]{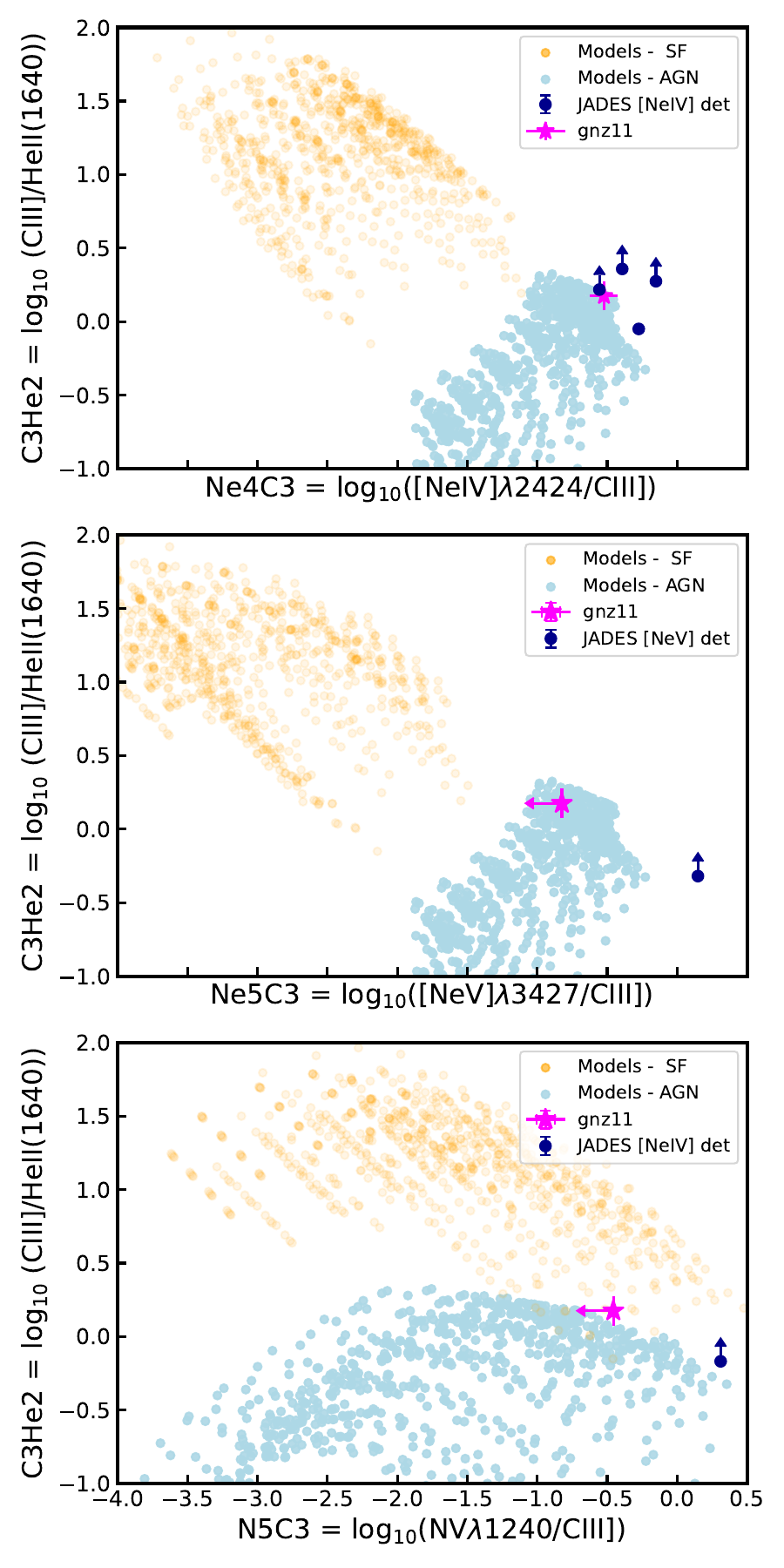}
   \caption{Emission line diagnostics plot for UV high ionisation lines. - Top panel: C3He2 vs \NeIVl2424/\CIII; Middle Panel: C3He2 vs \NeVl3427/\CIII diagnostic diagram; Bottom panel: C3He2 vs \NVl1240/\CIII. We plot our objects detected in high ionisation lines as blue points. The yellow and light blue points show star forming and AGN from photo-ionisation models from \citet{Feltre16,Gutkin16} (see \S \ref{sec:cloudy}). We highlight the emission line ratios of GN-z11 \citep{Maiolino23gnz11} as the magenta star.
    }
   \label{fig:NeIV_lines}
\end{figure}

We detect \NeIVl2424 in five objects. We plot these sources in the top panel of Figure \ref{fig:NeIV_lines}, showing C3He2 vs \NeIVl2424/\CIIIs as blue points, together with the photo-ionisation models of \citet{Gutkin16} and \citet{Feltre16} as yellow and blue points, respectively. All of these four sources have extremely high log$_{10}$(\NeIVl2424/\CIIIs) ratio ($>-0.6$) indicating that these sources are ionised by AGN. The fifth source (ID JADES-NS-GS-1000626) is poorly constrained in both \CIIIs and \HeIIl1640, and therefore we do not show it on this diagram.

We detect \NeVl3427 and NV$\lambda$1240 in the objects JADES-NS-GS-10013609 and JADES-NS-GS-00021842, respectively. In both cases, the detections are secured at SNR$\sim$4.5. However, the objects do not have solid detections of either \HeIIl1640 or \CIIIs. The non-detections of \CIIIs put a lower limit on the log(\NeVl3420/\CIIIs) or log(\NVl1240/\CIIIs) of $>$0. According to \citet{Feltre16}, this lower limit is on the border of what can be reasonably predicted by AGN photo-ionisation models. The \NVl1240 detection can be potentially explained by star formation, assuming a very high logU=-0.5, which is inconsistent with that measured from \OIIs and \OIII\ (logU=-2). As such, we identify JADES-NS-GS-00021842 as an AGN. It is necessary to observe these objects with deep rest-frame UV spectroscopy to help constrain the next geeneration of photo-ionisation models.

\section{Discussion}\label{sec:discussion}

In the previous section, we presented the selection of AGN in the JADES HST Deep survey. Overall, we selected 41 AGN candidates, from at least one of the different selection methods investigated in Section 3, in our parent sample of 209 galaxies (defined in \S \ref{sec:selection}) in the redshift range of 1.4--9.4. The final fraction of type-2 AGN in the galaxy population is up to 20\%. In this section, we discuss our results and their implications. In \S \ref{sec:stacking} we perform spectral stacking to find average emission line properties of AGN and star-forming galaxies, in  \S \ref{sec:AGN_sel_discussion} we compare the different selection methods used in this work, in \S \ref{sec:AGN_lum} we investigated the AGN bolometric luminosities, in \S \ref{sec:galaxy_prop} we compare the SFR and stellar masses of AGN to those of star-forming galaxies and finally in \S \ref{sec:UV_lum_fce} we discuss the contribution of AGN host galaxies to the UV luminosity function.

\subsection{Stacking}\label{sec:stacking}

In order to get the average emission line properties of our star-forming and AGN samples, we stacked the R1000 grating spectra for each of the emission lines used in our emission line diagnostics. Although the PRISM observations are deeper, the vastly varying spectral resolution as a function of wavelength makes any stacking efforts challenging. Furthermore, some of the emission lines we are interested in (such as [NII], \HeII1640) would be blended with other emission lines. Also, we only stack spectra from the 1210 program, since the 3215 observations lack band-2 observations (see discussion in \S \ref{sec:AGN_sel_discussion}). As such, this would bias the stack against redshifts where the key emission lines are redshifted to band-2. 

We stack the AGN hosts and star-forming galaxies in two separate stacks: a rest-frame UV emission line stack (\CIV, HeII$\lambda$1640 and \CIII) and a rest-frame optical one (HeII$\lambda$4685, \Hb, \OIII, \Ha, \NIIs). For each of these two cases, we stack all sources with available R1000 coverage of these emission lines. Overall, we stack 9 and 20 AGN, and 31 and 55 star-forming galaxies (i.e. the parent sample type-1 and type-2 AGN) in the UV and optical emission line sets, respectively, using only the 1210 program. As shown with type-1 AGN, some objects can be low luminosity AGN that \textbf{are} not selected using any of our diagnostics. As such, the star-forming galaxy sample can be contaminated by unidentified AGN. 

The stacking was performed by first shifting all spectra to the rest-frame and rebinning them to a common wavelength grid using Python's \texttt{spectres} package \citep{Carnall_spec}. We then model and subtract the continuum with a single power law. This is an appropriate model for the continuum, as we are only fitting a narrow wavelength range, and the continuum is poorly detected in the R1000 spectra. The individual spectra are weighted using two separate weighting schemes: a) 1/rms$^2$ weights; and b) 1/(rms$^2\times$ F$_{\rm[OIII]}$) or 1/(rms$^2\times$ F$_{\rm CIII]}$) weights for the UV line stacking) weights. However, given the large range of redshifts and galaxy luminosities, we found that the final stacked spectra are dominated by bright targets when we do not normalise by line fluxes. Hence, we will use the stacked spectra weighted by both the noise and the line fluxes (\OIIIs or \CIII), which approximates a median stacked spectrum.

The measured fluxes and their uncertainties for the HeII$\lambda$4685, \Hb, \OIII, \Ha, \NIIs and UV emission lines: CIV, HeII$\lambda$1640 and \CIII$\lambda$1907,1909 along with the number of objects in the stack and their median redshift are summarised in the Table \ref{table:stacks}. The final stacked spectra and the best fits are shown in the Appendix in Figures \ref{fig:stacking_agn} and \ref{fig:stacking_sf}. We detect all emission lines across all samples at $>3\sigma$, except for \HeII$\lambda4686$ in the star-forming galaxy sample. The detection of \NIIs seems like a contradiction compared to \citet{Cameron23}, however, this is purely due to the inclusion of galaxies with z<4, while \citet{Cameron23} restricted their sample to $z>5.5$.

We show the line ratios from the stacked spectra in all diagnostic diagrams for the AGN and star-forming samples as magenta and cyan squares in Figures \ref{fig:BPT_full}, \ref{fig:HeII_full} and \ref{fig:UV_lines}.

The stacked spectra on the N2-BPT (see top panel of Figure \ref{fig:BPT_full}) show that AGN at high-z are \NIIs weak and have a \OIII/\Hbs ratio very similar to star-forming galaxies, making the selection of AGN at high-z impossible using this method. Furthermore, the AGN host galaxies have the same \SII/\Has ratios as star-forming galaxies (see bottom panel of Figure \ref{fig:BPT_full}), further showing the difficulty of using this diagram to select AGN at high-z. 

We do not detect the \HeIIl4686 in the star-forming galaxies, and therefore we place a 3$\sigma$ upper limit on the flux. The \HeIIl4686/\Hb \ vs \NII/\Has diagram shows a separation of star-forming and AGN galaxies as expected based on the study by \citet{Shirazi12} that focused on SDSS galaxies. As such, the  \HeIIl4686  line is an ideal tracer of AGN activity; however the  \HeIIl4686  is a factor of 7 fainter than  \HeIIl1640, making it extremely difficult to detect, even with JWST. 

We detect \HeIIl1640 in both star-forming and AGN-dominated galaxies, although the flux of \HeIIl1640 is three times higher in AGN compared to star-forming galaxies. Furthermore, the \CIVs emission line is a factor of eight brighter in AGN than in the star-forming galaxies. This indicates a much higher and harder ionisation field in AGN host galaxies, boosting the high ionisation lines. However, the stacked spectrum of star-forming galaxies shows \HeIIl1640, with the star-forming stack in the border of AGN and star-forming galaxies on the \CIII/\CIV \,vs \CIII/\HeIIl1640 diagnostic plot (see Figure \ref{fig:UV_lines}). This can be due to some low luminosity AGN in the star-forming sample that do not have individual \HeIIl1640 detections, or because star-forming galaxies at high-z are creating more hard ionising photons than previously modelled.

\begin{table}
 \centering
   \caption{ Results of the stacking analysis for UV and optical emission lines of star-forming and AGN samples.}
 \begin{tabular}{lcccc}
  \hline
  Sample & AGN & AGN  & SF & SF \\
         & Optical & UV & Optical & UV \\
  \hline
  N & 20 & 9 & 55 &31 \\
  z$_{\rm median}$ & 3.87 & 5.93 & 4.77 & 5.91 \\
  \\
  Fluxes &($\times$10$^{-20}$ &erg s$^{-1}$ cm$^{-2}$) \\
  \hline
  \Ha   & 114.5$^{+1.7}_{-1.6}$ & - &  66.5$^{+0.8}_{-0.8}$ & -\\
  \NII  & 16.6$^{+1.0}_{-1.1}$ & - & 2.5$^{+0.4}_{-0.4}$& -\\
  \SII  & 4.5$^{+2.0}_{-1.7}$ & - & 1.6$^{+0.5}_{-0.5}$& -\\
  \OIIIs& 222.2$^{+1.9}_{-2.0}$ & - & 113.3$^{+0.8}_{-0.8}$ & -\\
  \Hb   & 48.8$^{+1.5}_{-1.5}$ & - & 24.1$^{+0.6}_{-0.7}$ & -\\
  \HeII$\lambda$4686 & 6.6$^{+1.2}_{-1.1}$  & - & $<10.5$ & -\\
  \hline
  \CIII & - & 20.7$^{+3.8}_{-3.7}$& -& 16.5$^{+1.9}_{-1.8}$\\
  \CIV  & - & 59.0$^{+4.7}_{-4.6}$& -& 6.1$^{+1.7}_{-1.5 }$ \\
  \HeII$\lambda$1640  & - & 18.6$^{+3.3}_{-3.3}$ & -& 4.2$^{+1.4}_{-1.3}$  \\
  \hline
 \end{tabular}
  \label{table:stacks}
\end{table}

\subsection{Comparing AGN selection methods}\label{sec:AGN_sel_discussion}

\begin{figure}
    \includegraphics[width=0.99\columnwidth]{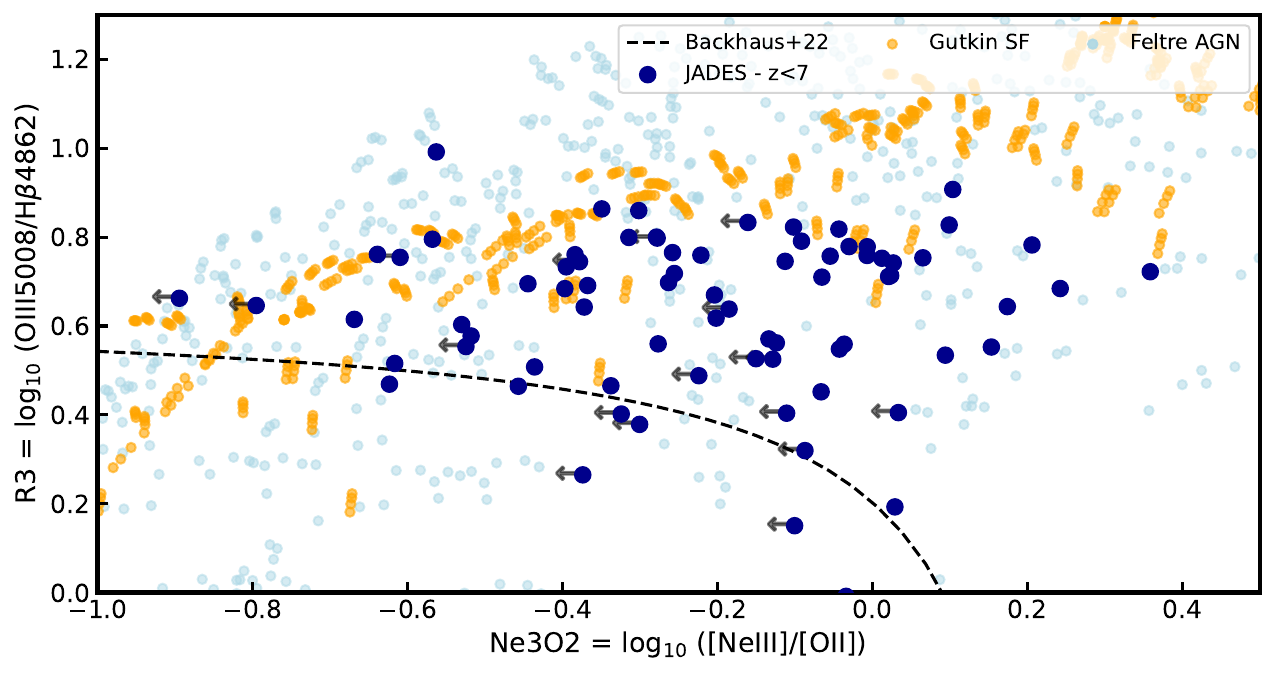}
   \caption{ \OIIIs/H$\beta$ vs [NeIII]/[OII] diagram ("OHNO"). The black dashed lines show the boundary between AGN and star forming galaxies from \citet{backhaus22}. The orange and light blue points show star forming and AGN photoionisation models from \citet{Feltre16} and \citet{Gutkin16}, respectively. There is a significant overlap between AGN and star forming galaxies in this diagram, making it unreliable to select AGN. 
    }
   \label{fig:NeIII_OII}
\end{figure}

Regardless of the emission diagnostic used to identify AGN (emission lines, X-ray observations, mid-infra-red selection), the selection method is reliant on finding emission that cannot be explained by star-formation processes. Objects not selected as AGN, the star-forming galaxies, can still have a significant amount of AGN activity, but it is not dominating the total galaxy emission. Therefore, AGN selection most likely provides an lower limit on the total number of AGN, as we are likely missing low luminosity active black holes, outshone by their host galaxy.

There is little overlap between the individual emission line ratios used to find AGN in this work, with only five objects being selected in more than one diagnostic. This can be explained by the different emission line diagnostics being sensitive to AGN identification at different regimes (e.g. redshift and metallicity of the ISM). As discussed above, the N2-BPT method becomes increasingly unreliable towards high redshift as galaxies become more metal-poor and have higher ionisation parameters. Although the highest redshift AGN selected by this method is at z=5.135 (JADES-NS-GS-00009452; 12+log(O/H)= 7.92), the bulk of the AGN selected by this method are below z=2.3. 

\citet{Topping2020, Runco2021} investigated the effect of stellar population age and metallicity on the BPT and S2-VO87 diagrams. These works have found that a population of young, metal-poor stellar populations can lie above the \citet{Kewley01} line. These low S2 ratio targets would be excluded from selection with the new definition of the demarcation lines and the AGN selected in the S2-VO87 diagram are above the points seen in both \citet{Topping2020, Runco2021} \citep[see also][]{Strom2017}. However, many of our selected AGN are very close to the demarcation line, making their selection tentative. As a result, we select any AGN selected within a distance of 0.1 dex from the demarcation as tentative, and we mark this with asterisks in Table \ref{table:Sample}. We note that our conclusions do not change whether we include or exclude these tentative AGN in our final sample.   

In both sets of observations (1210 and 3215) we select more AGN using the S2-VO87 diagram (with \SIIs emission line) than the classical BPT diagram (using \NIIs emission line). The \SIIs doublet is not blended with \Has in the PRISM observations at $z<5$, and hence we can use the deeper PRISM observations to constrain \SIIs doublet. On the contrary, the \NIIs doublet is blended with \Has in the PRISM observations, and hence we require the shallower R1000 JADES observations to put constraints on it. This is especially taxing for the 3215 observations, which were designed to observe galaxies at z$>6$ and hence do not have any R1000 Band 2 observations. This results in no constraints on the \NIIs doublet for z = 1.8--3.6 in the 3215 program, and hence low detection of AGN on the BPT diagram in this program. 

Luckily, as we push our AGN selection to higher redshifts, many useful UV lines (\NeIVl$\lambda$2424, \NeVl3427, \NVl1240, \HeIIl1640) are redshifted into the wavelength range of NIRSpec, and then to a higher sensitivity range of NIRSpec ($\lambda>1.2 \mu$m). As a result, the majority of the AGN selected above z=3 are based on \NeIVl$\lambda$2424, \NeVl3427, \NVl1240, \HeIIl1640. Still, these essential emission lines for identifying AGN activity at high redshift remain difficult to detect with JWST. Many of the emission lines such as \NVl1240, \HeIIl1640 are blended with nearby emission lines in the PRISM observations and require R1000 observations to deblend, which are shallower in JADES survey compared to the PRISM observations. Dedicated ultra-deep R1000 observations are required to detect these lines, even in moderate luminosity AGN. 

The HeII$\lambda$4686 emission line remains a robust diagnostic to detect AGN across large redshift and metallicity ranges. However, it is intrinsically fainter by a factor $\sim$10 compared to the HeII$\lambda$1640, resulting in it being a challenge to detect even in deep JWST spectra. However, if detected at high redshift, it is an unambiguous tracer of AGN activity at high redshift. 

It is important to point out that these high-ionization lines do not trace AGN activity as such, but a hard ionising radiation, with which AGN activity is the most likely source in galaxy evolution. However, other sources can also produce \HeII emission such as Wolf-Rayet stars, X-ray binaries, and some more exotic star-formation processes (e.g. very top-heavy IMF; \citealt{Thuan05, Kehrig15, Schaerer19, Umeda22}). However, the WR stars produce very broad \HeII forming so-called "blue bump" (see \citealt{Brinchmann08}) features not observed in any of our sources. Furthermore, \citet{Saxena20} indeed reproduced their detections of \HeII in the VANDELS survey using implementations of binary stars using BPASS models. However, the AGN and SF galaxies from the same VANDELS survey \citep{Mascia23} show the separation of the SF and AGN host galaxies predicted by the \citet{Gutkin16} and \citet{Feltre16} models. \citet{Nakajima22} models, which we use as a comparison of our data, also use BPASS to implement hard ionising photons from binary stars. Meanwhile, high-luminosity AGN at high redshift, such as type-1 AGN at z$\sim$5.5 from \citet{Ubler23} do show features of relatively weak \HeIIl4686 (log$_{10}$(\HeIIl4686/\Hbs)$\sim -1.2$), showing that even objects with relatively weak HeII can be AGN. Finally, \citet{Cameron23_9422} identified one of our AGN (ID 9422) to be dominated by a nebular continuum. This will be addressed in a separate paper in more detail (Tacchella et al. in prep).

\citet{Larson23} investigated the use of "OHNO" diagram (\OIIIs/H$\beta$ vs [NeIII]/[OII]) to replace the N2-BPT diagram in selecting AGN. We plot the Cloudy photoionisation models in Figure \ref{fig:NeIII_OII} with JADES galaxies. In this diagram there is a significant overlap between AGN and star-forming galaxies. The one galaxy with \OIIIs/H$\beta$>0.95 lies in the region only populated by AGN Cloudy models is 16745 at $z=5.56$. Overall, in agreement with \citet{Larson23},  we confirm that it is difficult to differentiate between AGN and metal-poor high ionisation star-forming HII regions in the OHNO diagram. For this reason, we need to rely on higher ionisation lines. 
 
Finally, \citet{Maiolino23JADES} recently published a sample of type-1 AGN observed with two sources overlapping with our observations. We detected only one as an AGN based on emission line diagnostics (JADES-NS-GS-0008083). Although it was selected by detection of both HeII$\lambda$4686 and \NeIV, the other AGN (with log(L$_{\rm Bol}$/(ergs s$^{-1}$)=44.3) is not selected in any of the narrow emission line diagnostics. As already discussed above, it is important to stress that emission line diagnostics miss AGN at high redshift. This can be attributed to 1) the physical properties of high-z AGN being either different from local templates (especially because of low metallicity); 2) these AGN not properly modelled by photoionization models (hence not predicted by the diagnostics); or 3) the emission from these sources is dominated by star-formation in the host galaxy (see Silcock et al. in prep).


\subsection{AGN fraction with deeper and shallower data }\label{sec:tier_comp}

Across the two observational programs, we selected 28 and 14 AGN from the parent samples of 110 and 99 galaxies, respectively. This corresponds to an AGN fraction of 24$\pm$5 \% and  14$\pm$4 \%. We note that the Poisson statistic most likely underestimates the uncertainties on the AGN fraction as the fraction of recovered AGN is likely dominated by the uncertainties in the selection criteria using the emission line diagnostics. More specifically, the final selection of 42 AGN across 1210 and 3215 programmes includes 10 {\it candidate} AGN; this suggests that the more reliable number of AGN in the sample is 30, and that there is a $\sim$25\% uncertainty on the AGN fraction arising from the diagnostics uncertainty. As such, the final AGN fraction across 1210 and 3215 programmes is $\sim$20$\pm$5 \%. 
We stress that this is the fraction of AGN within the JADES parent sample, and may not be representative of the AGN fraction in other samples that are characterized by other selection criteria.
We however note that the estimated AGN fraction of 20$\pm$5\% is consistent with estimates in other surveys such as CEERS \citep{Mazzolari24b})

\begin{figure}
    \includegraphics[width=0.9\columnwidth]{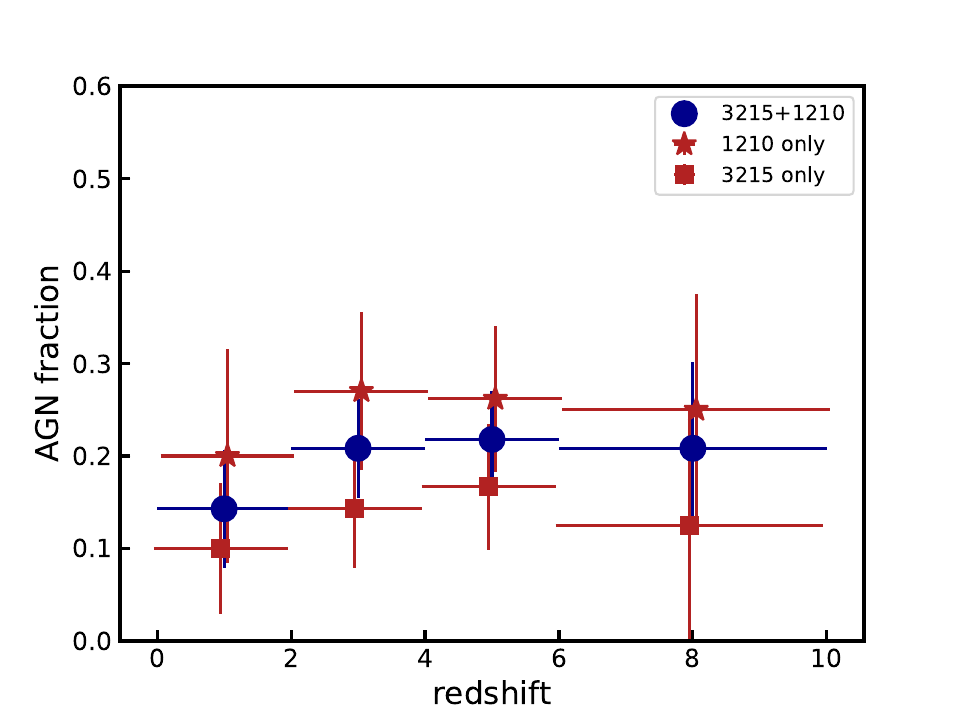}
   \caption{ Fraction of AGN as a function of redshift in our survey. The blue points show the fraction resulting from the combined programs (1210+3215) while the red stars and squares show separately the 1210 and 3215, respectively. The red points are offset by 0.05 in the x-axis for better visibility. We see no significant evolution of the AGN fraction as a function of redshift. 
    }
   \label{fig:AGN_fraction}
\end{figure}

In Figure \ref{fig:AGN_fraction}, we show the AGN fraction as a function of redshift. We plot the AGN fraction from 1210 and 3215 programs as red stars and squares and the combined dataset as blue points. The AGN fraction is consistent with being constant across all redshifts within 1$\sigma$.

The fraction of AGN between the two tiers is consistent within 1.5$\sigma$ of each other. Although this difference is statistically not significant, it is worth discussing the apparent decrease of AGN in the 3215 program, despite this program being a factor of 1.5-3$\times$ deeper than 1210 program. In the 3215 observations, only two AGN host galaxies were selected based on \HeIIl4686, compared to nine in the 1210 program. This is primarily due to the lack of Band 2 R1000 observations in 3215, which results in no constraints of \HeIIl4686 for redshifts 2.8--5.4. As discussed above, this similar issue is also plaguing the BPT selection using the \NIIs emission line, and hence we do not select any AGN using the diagnostics in BPT either. 

The Band 1 R1000 observations, which are key to observing rest-frame UV emission lines such as \CIII, \HeIIl1640 and \CIVs doublet, are only a factor of $\sim$1.4 deeper (twice the exposure time) in 3215 than in 1210. Unfortunately, this improvement in sensitivity is not enough to constrain the \HeIIl1640 in regular SF galaxies or less extreme AGN. We would thus require deeper rest-frame UV observations to constrain the AGN and star-forming populations (over 50 hours with JWST/NIRSpec).

\subsection{AGN luminosities}\label{sec:AGN_lum}

\begin{figure*}
    \includegraphics[width=0.99\textwidth]{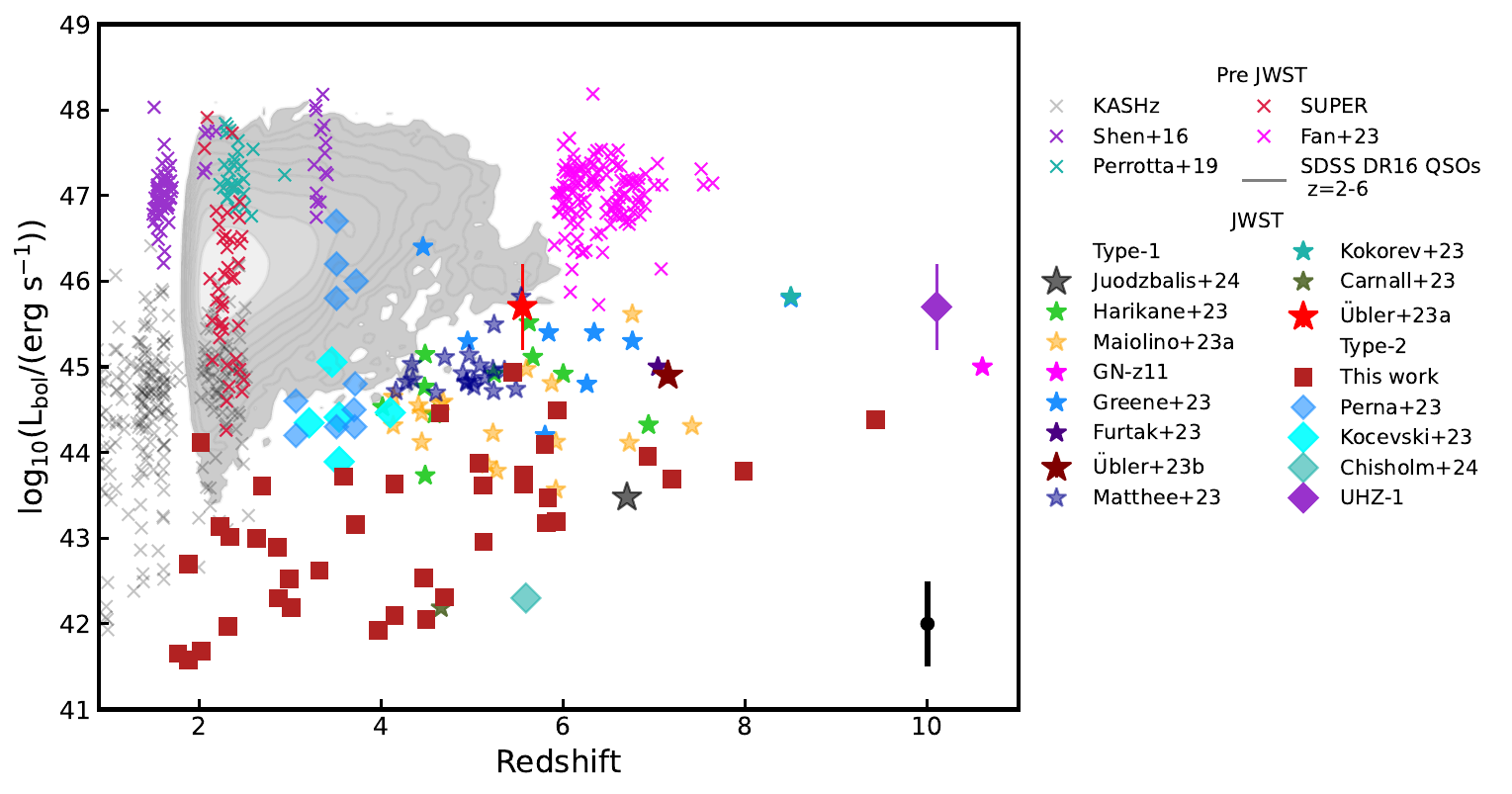}
   \caption{ Bolometric luminosity vs redshift for our objects, plotted as red diamonds. Our bolometric luminosities are estimated from dust-corrected narrow emission lines and should be treated as upper limits. The black point in the bottom right corner shows the systematic uncertainty on the bolometric luminosity for Type-2 AGN. Systematic uncertainties are dominating our measurements.  We compare our sample with previous JWST type-1 AGN \citep[various coloured stars and diamonds][]{Bogdan23, Carnall23b,Kocevski23, Kokorev23, furtak23,Goulding23, Maiolino23JADES, Maiolino23gnz11, Matthee23, Ubler23, Chisholm24}, AGN from the KASHz and SUPER surveys at Cosmic Noon \citep[grey and red crosses][]{Harrison16, Kakkad20} and QSOs samples across redshifts: SDSS QSOs \citep[z=2-6, grey shaded region;][]{Wu22}, extremely red quasars \citep[blue-green crosses;][]{Perrotta19}, blue QSOs \citep[purple crosses;][]{Shen16} and compilation of EoR QSOs \citep[magenta crosses;][]{Fan23}.
    }
   \label{fig:Lbol}
\end{figure*}

The bolometric luminosity is one of the key properties of AGN and can be easily estimated through the luminosity of the X-ray, BLR or the UV continuum emission from the accretion disc \citep[e.g.][]{Stern12, Netzer19,Duras20,Saccheo23}. However, as our sources were not selected as Type-1 AGN or X-ray detections, but based on their narrow emission line properties, hence, we do not have the BLR properties of these AGN, nor we have their UV or X-ray flux. As a result, we are forced to use the narrow line emission lines to estimate the bolometric luminosity. We calculate the bolometric luminosities (L$_{\rm bol}$) of our sample from the narrow line dust corrected fluxes of \Hbs, \OIIIs and \CIIIs, using the new calibrations by Hirschmann et al. (in prep), using the A$_{\rm V}$ estimated from BEAGLE fitting (see \S \ref{sec:SED_fitting}). Compared to the calibration from \citet{Netzer09}, our calibration has an additional quadratic term for the emission line luminosity, while \citet{Netzer09} depends only linearly on the line luminosity. The estimated L$_{\rm bol}$ from all three emission lines agree within 2$\sigma$, confirming the validity of the calibration. However, the values are systematically lower by 0.8 dex compared to the values obtained when using \citet{Netzer09} calibrations. 

We present the estimated bolometric luminosities in Figure \ref{fig:Lbol} as a function of redshift, together with other AGN from the literature (both from JWST spectroscopic studies and from previous, non-JWST surveys). However, these luminosities should be considered as upper limits, as they assume that the \Hbs and \OIIIs or \CIIIs emission is dominated by the narrow line emission from the AGN, with no contribution from star formation. This is not necessarily true, as AGN are hosted in star-forming galaxies (see \S \ref{sec:galaxy_prop}). The estimated bolometric luminosities are reported in Table \ref{table:Sample}.

As illustrated in Figure \ref{fig:Lbol}, pre-JWST studies used large optical and NIR surveys to identify sources dominated by bright rest-frame optical and UV emission, selecting primarily QSOs, or intermediate luminosity AGN at z$<$3. Specifically, in Figure \ref{fig:Lbol}, we show a compilation of QSOs by \citet{Fan23} as green, blue and magenta crosses and X-ray AGN from the KASHz and SUPER surveys \citep{Harrison16,Circosta18,Kakkad20}, and selection of red and blue QSOs from Cosmic Noon from \citet{Shen16, Perrotta19}. We also show quasars from SDSS DR16 \citep{Wu22} as shaded contours. 
Since the launch of JWST, there have been a number of studies searching for AGN \citep[plotted in Figure \ref{fig:Lbol} as various coloured diamonds and stars, ][]{Bogdan23, Carnall23b,Kocevski23, Kokorev23, furtak23,Goulding23, Maiolino23JADES, Maiolino23gnz11, Matthee23, Ubler23, Chisholm24}. Our AGN sample has generally similar luminosities to those selected as type-1 AGN with JWST spectroscopic surveys.

With the new capabilities of JWST, we are now probing, at z$>$3, AGN that are 2-3 orders of magnitude less luminous than previous surveys, allowing us to understand the broader demographics of the AGN population in the early Universe for the first time.

\subsection{Host Galaxy properties}\label{sec:galaxy_prop}

One of the key questions of studying AGN host galaxies is investigating their star-formation properties. Until the launch of JWST, these studies have mostly focused on $z<3$ moderate luminosity AGN and $z>4$ quasars. These studies have shown that AGN host galaxies have SFR at, or just below, the level of the star-forming galaxies main sequence (SFMS) at Cosmic Noon when mass-matching the active and inactive samples \citep[$z\sim1-3$;][]{Santini12,Rosario13b,Vito14b,Mullaney15,Stanley15,Azadi15,Scholtz18, ForsterSchreiber19}.  With the launch of JWST, we can now select and measure the SFR of moderate luminosity AGN at high redshift. Additionally, unlike previous studies with JWST focusing on type-1 AGN, we do not suffer from the AGN contaminating the continuum emission in our spectra, and hence modelling the stellar SED of the host galaxy is less problematic (although the AGN can still contaminate the Balmer emission lines hence making them less reliable for the SFR estimation). In order to verify our SED fitting, we also used the SED fitting results from NIRCam photometry only performed by \citet{Simmonds24b} and we do not find any differences in our conclusions. 

\begin{figure}
    \includegraphics[width=0.95\columnwidth]{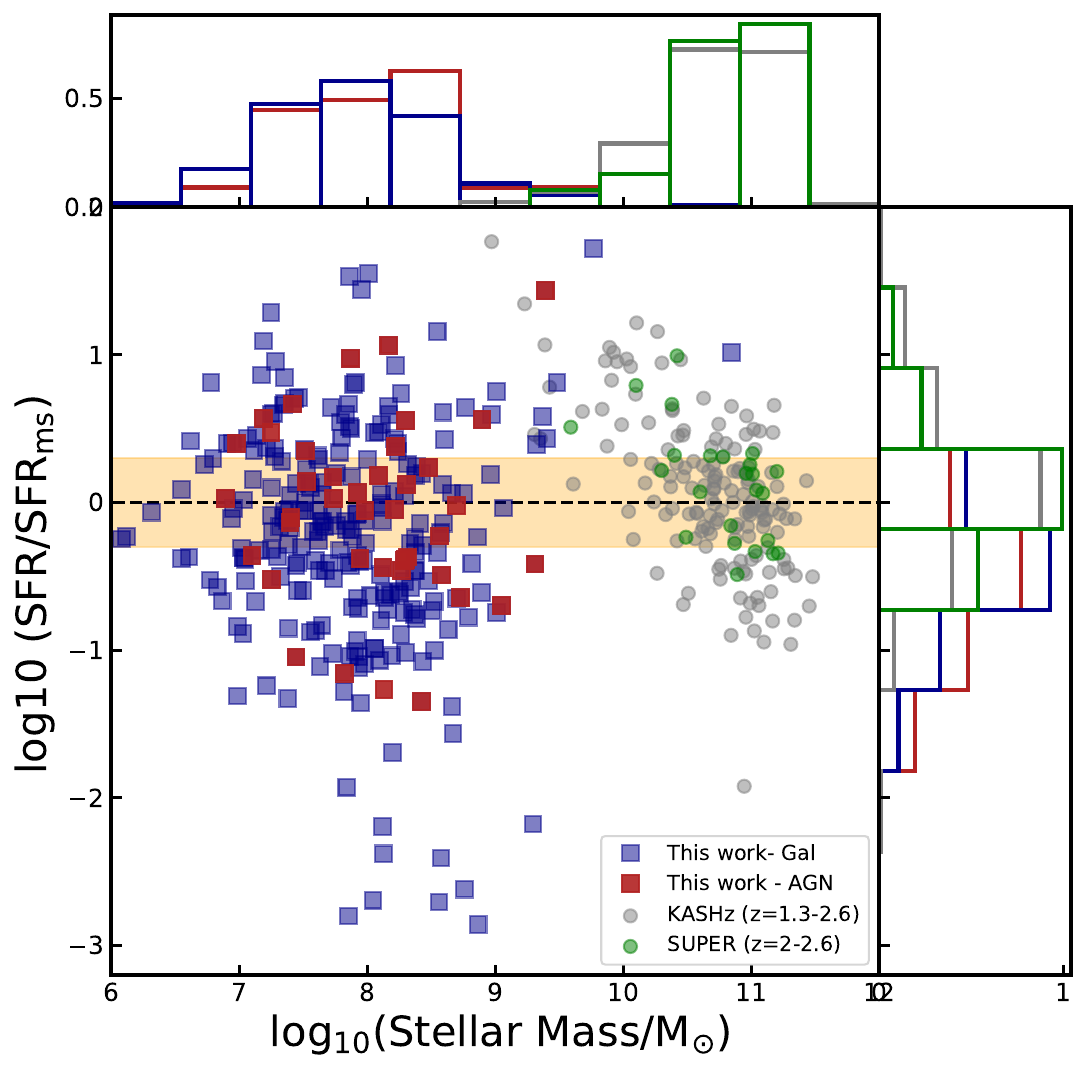}
    \includegraphics[width=0.95\columnwidth]{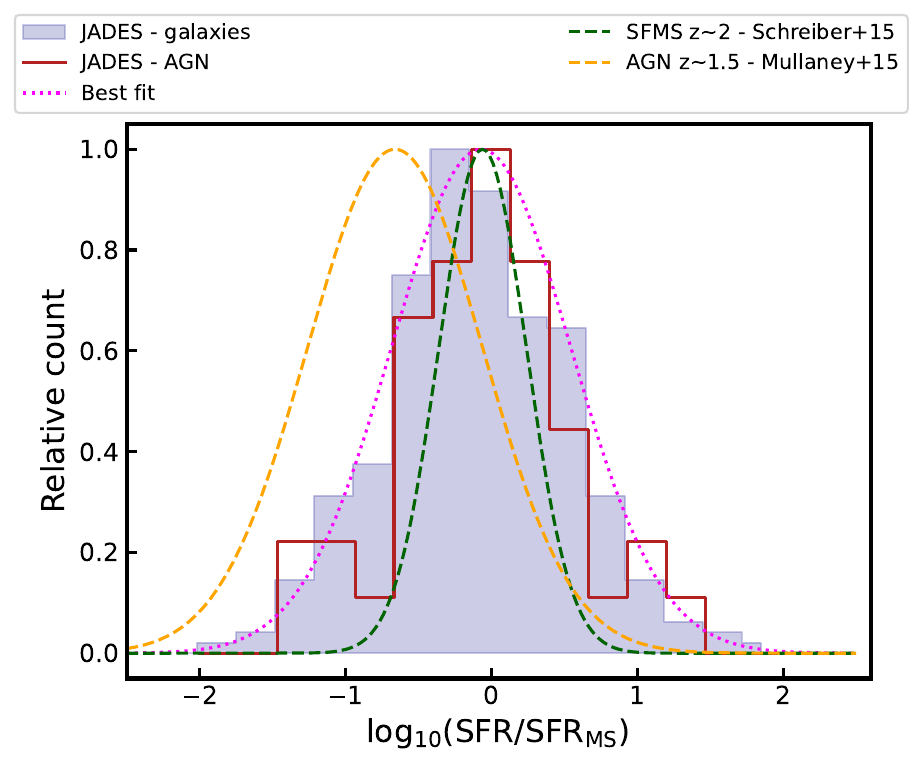}
   \caption{ Offset from the star-forming main sequence (SFMS) of our AGN host galaxies. Top panel: Offset from the SFMS against stellar mass for the JADES sample (blue points), our selected AGN (red points). We compare our sample to the previous AGN sample KASHz survey \citep[][grey points]{Harrison16} and SUPER survey \citep[][green points]{Circosta18}. We highlight the 1$\sigma$ scatter of SFMS by the yellow shaded region. Bottom panel: Distribution of SFR/SFR$_{\rm MS}$ of the AGN candidates selected in this work (red histogram). The blue histogram shows the data from this work, and the magenta dotted line shows the best fit to the distribution of offset from SFMS assuming it is log-normal. The green and orange dashed lines show the distribution for the SFMS \citep{Schreiber15} and AGN \citep{Mullaney15}, respectively. 
    }
   \label{fig:stellar_mass}
\end{figure}

In \S \ref{sec:SED_fitting}, we described the SED fitting approach adopted to derive the SFR and stellar masses. Here we further highlight that the SFR used in this work are averaged from the past 10 Myr. However, the results of our analysis do not change if we use SFR averaged over a different time span (e.g. 100 Myr). We use the star-forming main sequence (SFMS) prescription from \citet{Looser23SFH}, who estimated the SFMS based on 3 separate redshifts bins for sources from JADES with M$_{*}< 10^{9.3}$M$_{\odot}$. For AGN at Cosmic Noon, \citep{Harrison16, Circosta18} we use the SFMS prescription of \citet{Schreiber15} as it covers the redshift and stellar mass range of our AGN. It is worth mentioning that \citet{Schreiber15} includes a turnover at high stellar masses. 

In Figure \ref{fig:stellar_mass}, we show the offset from the star-forming main-sequence (SFR/SFR$_{\rm MS}$) against the stellar mass for AGN and star-forming galaxies in the JADES survey (red and blue points); KASHz AGN survey (grey points) and SUPER AGN survey (green points). For sources from the JADES survey, we use the star-forming main sequence defined in \citet{Looser23SFH} as it was defined within the redshift and mass range of our sample. The AGN host galaxies in our sample have the same stellar mass distributions as the star-forming galaxies in the JADES, with a median stellar mass of SF galaxies and AGN are $10^{7.9}$ and $10^{8}$ M$_{\odot}$, respectively. The presence of AGN at such low masses indicates that AGN activity is also important in galaxies with M$_{*}<10^{10}$, as previously suggested by at high redshift \citep{Koudmani21,Koudmani22} and low redshift \citep{Burke22,Mezcua23, Siudek23}.

The SFRs of our selected AGN are consistent with the galaxies (star-forming, quiescent and star-bursting alike) in the JADES sample. Performing K-S test (Kolmogorov–Smirnov test) on the distribution of offset from the Main-Sequence for JADES galaxies and AGN, we estimate a value of P=0.53, indicating that they are indeed originating from the same underlying distribution. 


To evaluate the difference in the star-formation properties of AGN in our sample, we investigate the distribution of SF$_{\rm offset}$. Following \citet{Mullaney15,Scholtz18, Bernhard19}, we fitted the distribution of SFR/SFR$_{\rm MS}$ assuming that it is log-normal, with mode and width as free parameters, using an MCMC method. The mode and width of the derived log-normal distribution are $0.07\pm 0.10$ and $0.61\pm0.07$, respectively. The mode of the distribution is consistent with that of SFMS (mode$_{\rm SF}$ = -0.07; \citealt{Schreiber15}), however, the width is 2$\sigma$ higher than that of SFMS (width of SFMS of 0.3 dex). Some of this difference may be attributed to the methods used to derive SFR \citep{Caplar2019}.  For example, the BEAGLE fitting accounts for physical variations that can lead to variation in the UV to SFR and H$\alpha$ to SFR relationships (see e.g. \citealt{Curtis-lake21}), while the UV and IR SFRs in \citet{Schreiber15} use simple conversions to SFR.

\citet{Mullaney15} have measured the distribution of SFR/SFR$_{\rm MS}$ for X-ray selected AGN at z$\sim$1.6, finding that the mode and the width of the distribution as $-0.36\pm0.07$ and $0.56\pm0.09$, respectively.
The disagreement between our results and \citet{Mullaney15} can be attributed to our selection of AGN as well as the fact that SFRs here should be considered upper limits. Furthermore, \citet{Mullaney15} investigated AGN in massive galaxies (M$_{*}> 10^{10}$M$_{\odot}$) at z=1-1.5. The majority of the high-z AGN were selected based on the detection of \HeIIl1640 or \HeIIl4686 which are more likely to be detected in brighter targets.

However, in this work, we do not have a sufficient number of sources to further split our AGN sample by redshift, stellar mass or AGN luminosity. This will be performed in future works with the full JADES sample. 

\subsection{Contribution of AGN host galaxies to the UV luminosity function}\label{sec:UV_lum_fce}

\begin{figure}
    \includegraphics[width=0.99\columnwidth]{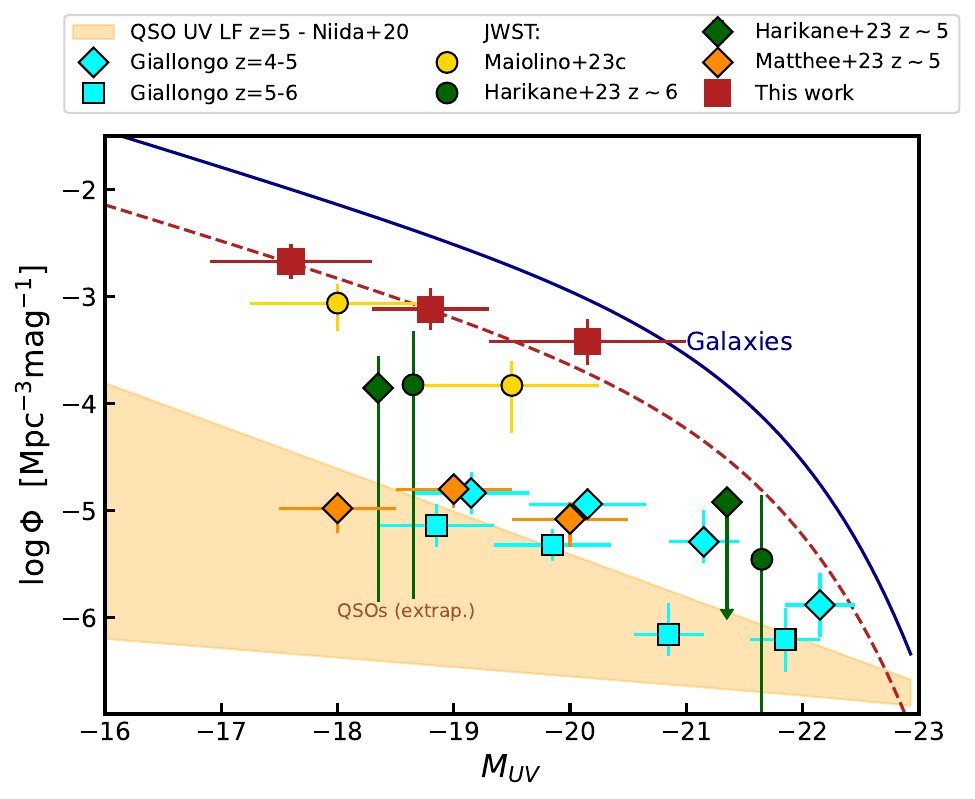}
   \caption{UV luminosity function of galaxies at z=5 from \citet{Bouwens21} (blue solid line) and inferred contribution of galaxies hosting type-2 AGN, inferred from the JADES survey (red points). Results from other surveys are also reported, as indicated in the legend. The dashed red line shows the galaxy luminosity function scaled downward by a factor of 5. The orange-shaded region shows the range of possible extrapolated luminosity functions for QSOs from \citet{Niida20}.
    }
   \label{fig:uv_lum}
\end{figure}

With this new selection of a large sample of Type-2 AGN, we can now investigate the contribution of AGN host galaxies to the UV luminosity function at high redshifts.
We explore the redshift range 4--6, as this is poorly explored in the literature before JWST and it is a redshift range in which we have good statistics.

Since the selection function for allocating targets in JADES is complex, it is hard to derive the volume density of AGN host galaxies as a function of the UV luminosity based on the number of AGN that we have identified. More importantly, we do not have high enough statistics to derive a proper luminosity function. Therefore, we make the simplified assumption that the spectroscopic selection function has not preferentially favoured or disfavoured galaxies hosting AGN. As discussed above, as the continuum luminosity is dominated by stellar light (and `normal' star forming galaxies at such high redshifts are also strong line emitters), the JADES spectroscopic sample should not be biased in favour or against AGN. Additionally, the JADES selection function, at least for its `HST-deep' tier (programme 1210), is fairly similar to the selection function used by \cite{Bouwens21} for inferring the UV luminosity function. Given these considerations,
we estimate the contribution of AGN host galaxies to the UV luminosity function simply as the fraction of AGN identified relative to the parent spectroscopically selected population, in UV luminosity bins.

We choose the parameterised luminosity function from \cite{Bouwens21} as the reference luminosity function, choosing a median redshift of our type-2 AGN at z=4--6 of z$_{\rm median}\sim5.5$ (14 objects in total).

Following the procedure discussed above, we show in Figure \ref{fig:uv_lum} (as red points) the UV luminosity function of the type-2 AGN host galaxies z$>4$ from the HST Deep of JADES survey. We split our sample into three separate bins, choosing the bins to allow at least three objects per bin, centred on $M_{UV}=$ -20.15, -18.8,  -17.6. The estimated contribution of the type-2 AGN host galaxies to the UV luminosity function is 33$\pm$4 \%, 18$\pm$4 \% and 20$\pm$4 \%, for the $M_{UV}=$ -20.15, -18.8,  -17.6 bins, respectively. We estimate the errors in the same ways as in \S~\ref{sec:tier_comp}. Given the limited number of points, we do not attempt to fit a functional form, as a small number statistics mean that the parameters are not adequately constrained. Instead, we simply show that the type 2 AGN host galaxies LF can match the \cite{Bouwens21} galaxy UV luminosity scaled by a factor of 5, as shown by the red dashed line. 

We also compare our results with other JWST type-1 AGN surveys: \cite{Harikane23}, \cite{Kocevski23}, \cite{Maiolino23JADES}, \cite{Matthee23}, as indicated in the legend. The cyan diamonds and squares show the luminosity function inferred by  \cite{Giallongo19} based on X-ray surveys. Given the uncertainties, our estimated type 2 AGN density is slightly higher (by a factor of $\sim$2) than the one estimated for type-1 AGN from the JADES and CEERS surveys \citep{Maiolino23JADES,Harikane23}.
We note that the contribution to the UV luminosity function is significantly higher than the estimate of \cite{Kocevski23}, but which was most likely suffering from a low number of statistics in the early JWST results. The contribution is also higher than found by \cite{Matthee23}, however, as discussed in \cite{Maiolino23JADES}, that study probes more luminous AGN, hence likely less abundant.

We also note that our estimated density of type-2 AGN is higher than found in the deep X-ray observations by \cite{Giallongo19}. As stated by \citet{Maiolino23JADES}, this suggests that the optical and UV emission line selection of AGN is picking a population of faint or X-ray deficient population of AGN at high-z. This has indeed been confirmed by the detailed analysis of the X-ray properties of this and other samples of JWST-discovered AGN, finding that they are extremely X-ray weak, even when stacked \citep{Maiolino2024X}. The origin of such X-ray weakness is still debated -- it is possibly due to a combination of heavy absorption and intrinsic weakness \citep{Maiolino2024X}.

The extrapolation of the QSO luminosity function at z$\sim$5.5 \citep[orange shaded region][]{Niida20} is 1--2 orders of magnitude lower than AGN luminosity function from JWST. Together with Figures \ref{fig:Lbol}, \ref{fig:stellar_mass} and \ref{fig:uv_lum}, this further illustrates that JWST is probing a new parameter space of low luminosity AGN activity at high-z redshift. 




\subsection{JADES-GS+53.11243-27.77461 - type-2 AGN at z$\sim$9.43}\label{sec:highz_agn}

\begin{figure}
    \includegraphics[width=0.99\columnwidth]{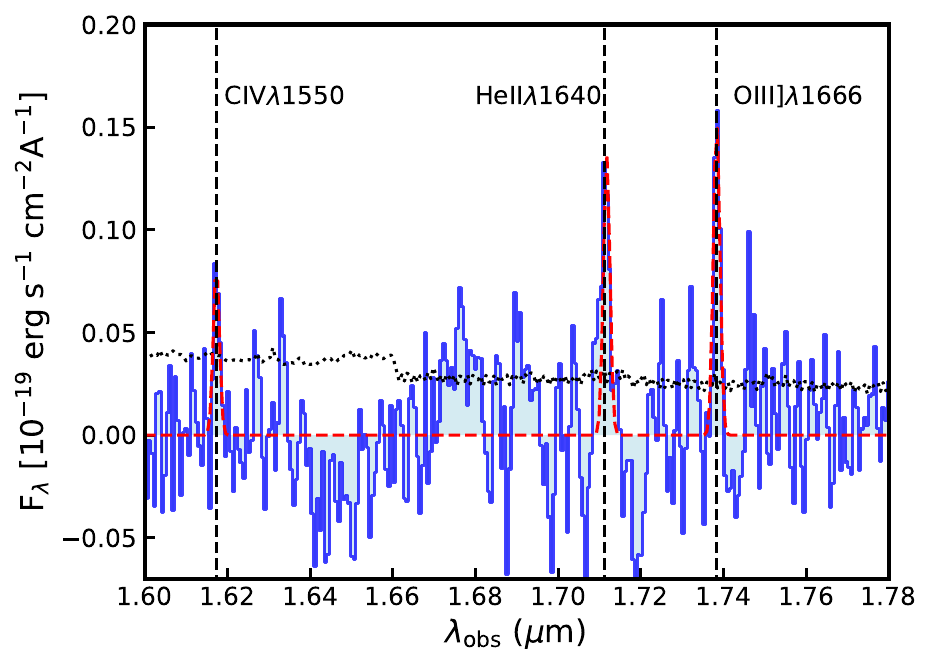}
   \caption{Detection of \CIVll1548,50, \HeIIl1640 and \OIIIl1666 in object ID 10058975 (JADES-GS+53.11243-27.77461) at z=9.43. The spectrum is a combined spectrum of the Band-1 and Band-2 data from program ID 1210 and Band-1 data from program ID 3215 to increase the SNR of the detection. \textbf{The black dotted line shows the flux uncertainties.} The blue line shows the continuum subtracted observed spectrum, while the red dashed line shows the best fit to the data.  
    }
   \label{fig:94HeII}
\end{figure}

The most distant type 2 AGN in our sample is ID 10058975 (JADES-GS+53.11243-27.77461), which was identified using two separate methods: 1) the detection of \NeIVl2422 UV emission line; 2) by identifying \HeIIl1640, \CIIIs and \CIVs emission line ratios diagnostic. \HeIIl1640 is redshifted to the $\sim$1.71 $\mu$m, which is covered by both Band-1 and Band-2 R1000 grating observations from 1210 observations and Band-1 observations in 3215 observations. We showed the final stacked spectrum from all three observations in Figure \ref{fig:94HeII}. The \HeIIl1640 emission line is detected in all three sets of observations, however, we stacked the three spectra for more robust detections (SNR=6.5). The \NeIVl2424 emission is detected at 4$\sigma$, and we show the spectrum in Figure \ref{fig:High_ion}, indicating the presence of strong ionising radiation. 

As this object is well detected in all gratings and PRISM observations, we were able to fit the full 1-5 $\mu$m spectrum with BEAGLE. We estimated the stellar mass of the AGN host galaxy to be $1.5 \times 10^{8}$ M$_{\odot}$ with an SFR of 6.6 M$_{\odot}$ yr$^{-1}$ (SFR/M$_{*}= 40$ Gyr$^{-1}$). The object has a UV magnitude of M$_{\rm UV}=-20$ and is most likely going through a starburst. However, as discussed above, the high SFR can be explained by the contamination of the emission lines by the AGN, artificially increasing the estimated SFR of the system.

Furthermore, this object has a strong \OIIIl4363 emission \citep[5.1$\sigma$][]{Laseter23}. This allowed the authors to constrain the electron temperature (T$_{e}$) of 18400$\pm$2100K and metallicity  12 + log(O/H)=$7.46\pm0.11$. Such high temperatures and strong auroral line are anomalous in HII regions, but seen in the NLR of AGN \citep[e.g.][]{Brinchmann23}. The full detailed analysis of the \OIIIl4363 and \OIIIl1661,66 is presented in Curti et al. (in prep.). 

The metallicity inferred from the direct method is 4$\sigma$ lower than the value inferred by BEAGLE. Despite the unambiguous detection of an AGN and exquisite data provided by NIRSpec we are still hampered in the characterization of NLR and host for integrated spectra such as this one.  More advanced SED fitting of these objects using e.g. BEAGLE-AGN \citep{Vidal-Garcia22} is outside the scope of this paper and will be further investigated by Silcock et al. (in prep). 

\section{Conclusions}\label{sec:conclusions}

In this work we have presented the identification of obscured, i.e. narrow-line (type-2), AGN candidates in the two deepest spectroscopic fields of the JADES survey in GOODS-S, using rest-frame optical and UV emission lines. We have investigated the presence of AGN ionisation, in narrow line emission, by using classical optical N2-R3, S2-VO87, N2-He2 diagnostic diagrams, as well diagrams exploiting UV emission lines: \CIII, \CIV, \HeIIl1640, \NeIVl2424, \NeVl3420 and \NVl1240.

Based on our analyses we find:

\begin{itemize}
    \item At z>3, the N2-BPT and S2-VO87 diagnostic diagrams ([OIII]5007/H$\beta$ vs [NII]6584/H$\alpha$, and vs [SII]6730/H$\alpha$, respectively) are no longer able to clearly distinguish photoionisation due to type-2 AGN and star-forming galaxies, because low metallicity, high-z AGN and star-forming galaxies occupy the same space on these diagrams (see Figure \ref{fig:BPT_full}).
    However, we redefine a conservative demarcation line between AGN and star-forming galaxies on the BPT and S2-VO87 diagrams to allow for the shift high redshift star-forming galaxies. We identify five and seventeen AGN host galaxies on the N2-BPT and S2-VO87 diagrams, respectively. We stress that this is a very conservative selection and that many more AGN are certainly present on these diagrams, mixed with star forming galaxies.

    \item 
    Using the \HeIIl4686/\Hbs vs \NII/\Has diagnostic diagram we selected eleven AGN and an additional six AGN using the \CIII/\CIV \,vs \CIII/HeII$\lambda$1640 diagram (see Figures \ref{fig:HeII_full} \& \ref{fig:UV_lines}). Interestingly, the only X-ray detected AGN in our sample, is located in the star-forming region of the BPT diagram, while is confirmed as AGN in the HeII diagnostic diagrams.

    \item We detected the high ionization transitions \NeIVl2424, \NeVl3420 or \NVl1240 in seven galaxies. The luminosities of these high ionisation lines compared to the \CIII\, emission line classify these objects as AGN hosts galaxies. 

    \item  In total we selected 28 AGN in the PID-1210 programme and 14 AGN in the PID-3215 programme, resulting in AGN fractions of $24\pm5$\% and $14\pm4$\%, respectively. Combining the two samples, we find 41 unique AGN, and an overall AGN fraction of $20\pm5$\% in the JADES survey. We investigated the evolution of AGN fraction as a function of redshift (see Figure \ref{fig:AGN_fraction}) and did not find evidence for significant evolution. 

    \item By stacking the AGN and star-forming galaxies' rest-frame UV and optical spectra we confirmed that both samples have similar emission line ratios in the rest-frame optical N2, S2 and R3 emission line ratios. However, the populations are easily distinguished using the \HeIIl4686 and \HeII1640 emission lines (see cyan and magenta points on Figures \ref{fig:BPT_full}, \ref{fig:HeII_full}, \ref{fig:UV_lines}). 

    \item  The estimated bolometric luminosities using narrow-emission lines (\OIII, \CIII and \Hb) are in the range of $6\times 10^{41}$--$5\times 10^{45}$ ergs s$^{-1}$ (see Figure \ref{fig:Lbol}). The selected AGN host galaxies have a median stellar mass of 10$^{7.9}$ M$_{\odot}$ consistent with the median stellar mass of inactive galaxies of 10$^{8.0}$ M$_{\odot}$. 
    
    \item We investigated the distance from the star-forming main sequence, log(SFR/SFR$_{\rm ms}$). Our AGN candidates have a mode and width of the distribution of log(SFR/SFR$_{\rm ms}$) of $-0.07\pm0.10$ and $0.61\pm0.07$, respectively. The mode of distribution is consistent with SF galaxies while the width of the distribution is a factor of two broader than for star-forming galaxies, consistent with previous results at Cosmic Noon. 
 
    \item We estimated the contribution of the AGN host galaxies to the UV luminosity function on the order of $\sim20$\%, with a slight (increasing) dependence on luminosity (see Figure \ref{fig:uv_lum}).
  
\end{itemize}

\section*{Acknowledgements}

We thank Anna Feltre for providing us with the updated Cloudy models for star-forming and AGN with expanded emission line coverage. We thank Michaela Hirschmann for the productive discussion to help with the selection based on UV emission lines. We would like to further thank Dominika Wylezalek and Eduardo Banados.
JS, RM, FDE, WB, XJ, TJL acknowledge ERC Advanced Grant 695671 “QUENCH” and support by the Science and Technology Facilities Council (STFC) and by the UKRI Frontier Research grant RISEandFALL.  RM  acknowledges support by the UKRI Frontier Research grant RISEandFALL as well as funding from a research professorship from the Royal Society.
SCa and GV acknowledge support by European Union’s HE ERC Starting Grant No. 101040227 - WINGS.
AJB, AJC, JC, AS \& GCJ acknowledge funding from the "FirstGalaxies" Advanced Grant from the European Research Council (ERC) under the European Union’s Horizon 2020 research and innovation programme (Grant agreement No. 789056).
ECL acknowledges support of an STFC Webb Fellowship (ST/W001438/1).
This research is supported in part by the Australian Research Council Centre of Excellence for All Sky Astrophysics in 3 Dimensions (ASTRO 3D), through project number CE170100013.
DJE is supported as a Simons Investigator and by JWST/NIRCam contract to the University of Arizona, NAS5-02015.
Funding for this research was provided by the Johns Hopkins University, Institute for Data Intensive Engineering and Science (IDIES).
MP acknowledges support from Grant PID2021-127718NB-I00 funded by the Spanish Ministry of Science and Innovation/State Agency of Research (MICIN/AEI/ 10.13039/501100011033). 
BER, KH, ZJ, JL, MR, FS, and CNAW acknowledge support from the NIRCam Science Team contract to the University of Arizona, NAS5-02015. 
WB acknowledges support by the Science and Technology Facilities Council (STFC).
BRP acknowledges support from the research project PID2021-127718NB-I00 of the Spanish Ministry of Science and Innovation/State Agency of Research (MICIN/AEI/ 10.13039/501100011033).
MSS acknowledges support by the Science and Technology Facilities Council (STFC) grant ST/V506709/1.
The research of CCW is supported by NOIRLab, which is managed by the Association of Universities for Research in Astronomy (AURA) under a cooperative agreement with the National Science Foundation.
H{\"U} gratefully acknowledges support by the Isaac Newton Trust and by the Kavli Foundation through a Newton-Kavli Junior Fellowship.
This work was performed using resources provided by the Cambridge Service for Data Driven Discovery (CSD3) operated by the University of Cambridge Research Computing Service (www.csd3.cam.ac.uk), provided by Dell EMC and Intel using Tier-2 funding from the Engineering and Physical Sciences Research Council (capital grant EP/T022159/1), and DiRAC funding from the Science and Technology Facilities Council (www.dirac.ac.uk).
 The authors acknowledge use of the lux supercomputer at UC Santa Cruz, funded by NSF MRI grant AST 1828315.

%
%

\bibliographystyle{aa}
\bibliography{mybib}

\begin{appendix}
    
\section{Stacked spectra}

\begin{figure*}
    \includegraphics[width=0.9\paperwidth]{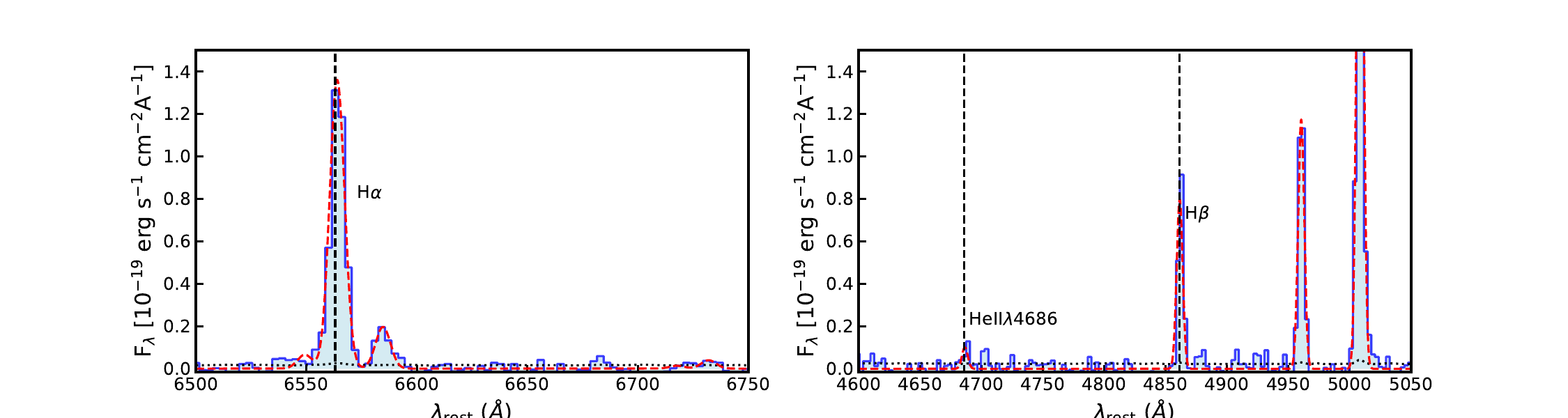}
    \includegraphics[width=0.9\paperwidth]{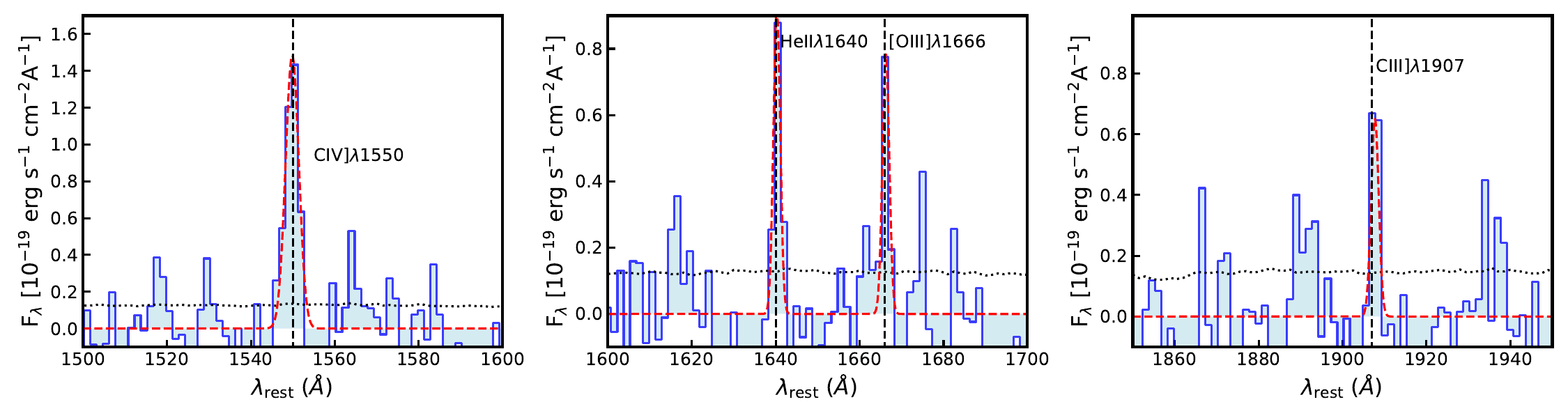}
   \caption{Resulting spectrum from the stacking analysis of AGN host galaxies. The blue line shows the continuum subtracted data, while the red dashed line indicates the best fit to the stacked spectrum. Top row: Stacked spectrum of \Ha, \NIIs and \SIIs (left panel) and \HeIIl4686, \Hbs and \OIIIl5008 (right panel). Bottom row: Stacked spectrum of \CIVs (left panel), \HeIIl1640 and \OIIIl1666 (middle panel), and \CIIIs (right panel).
    }
   \label{fig:stacking_agn}
\end{figure*}

\begin{figure*}
    \includegraphics[width=0.9\paperwidth]{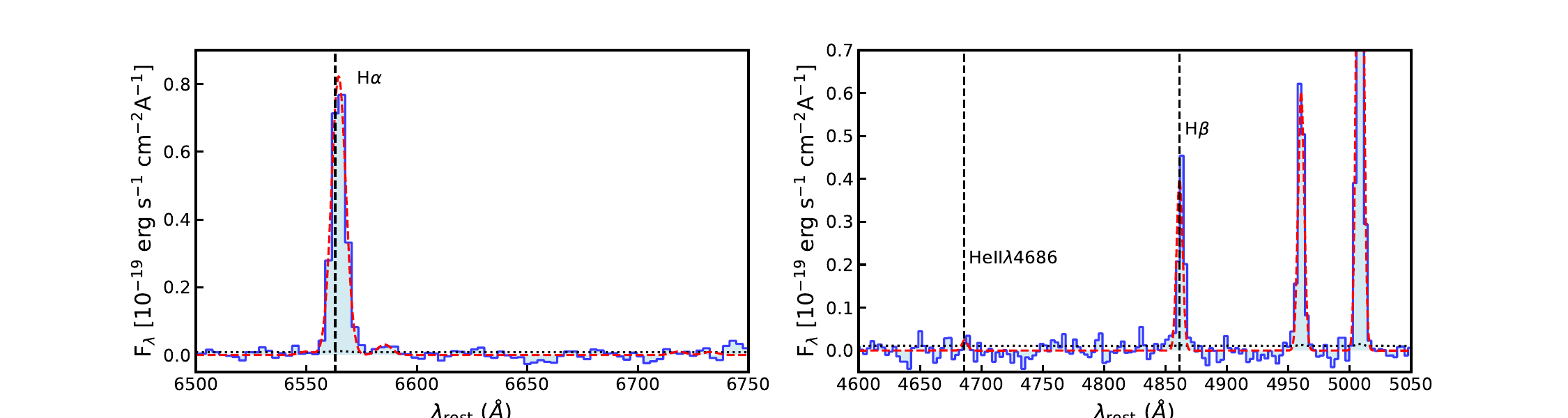}
    \includegraphics[width=0.9\paperwidth]{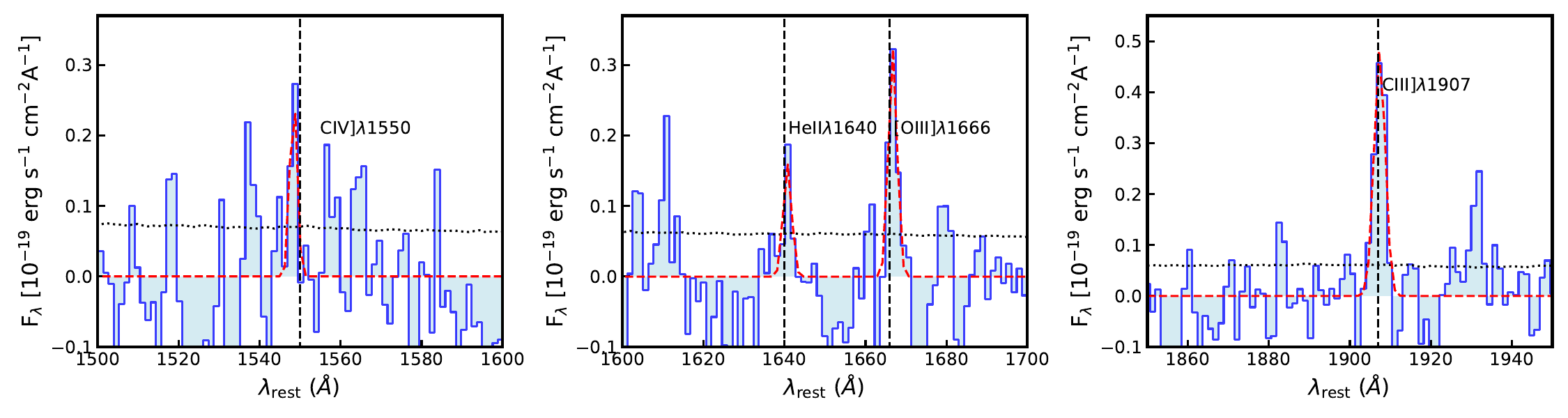}
   \caption{Resulting spectrum from the stacking analysis of star forming galaxies. The blue line shows the continuum subtracted data, while the red dashed line indicates the best fit to the stacked spectrum. Top row: Stacked spectrum of \Ha, \NIIs and \SIIs (left panel) and \HeIIl4686, \Hbs and \OIIIl5008 (right panel). Bottom row: Stacked spectrum of \CIVs (left panel), \HeIIl1640 and \OIIIl1666 (middle panel), and \CIIIs (right panel).
    }
   \label{fig:stacking_sf}
\end{figure*}

We show the stacked rest-frame UV and optical spectra in Figures \ref{fig:stacking_agn} and \ref{fig:stacking_sf} for the AGN and star forming galaxies.

\section{UV fits}

\begin{figure*}
     \centering
     \begin{subfigure}[b]{0.285\paperwidth}
         \centering
         \includegraphics[width=0.28\paperwidth]{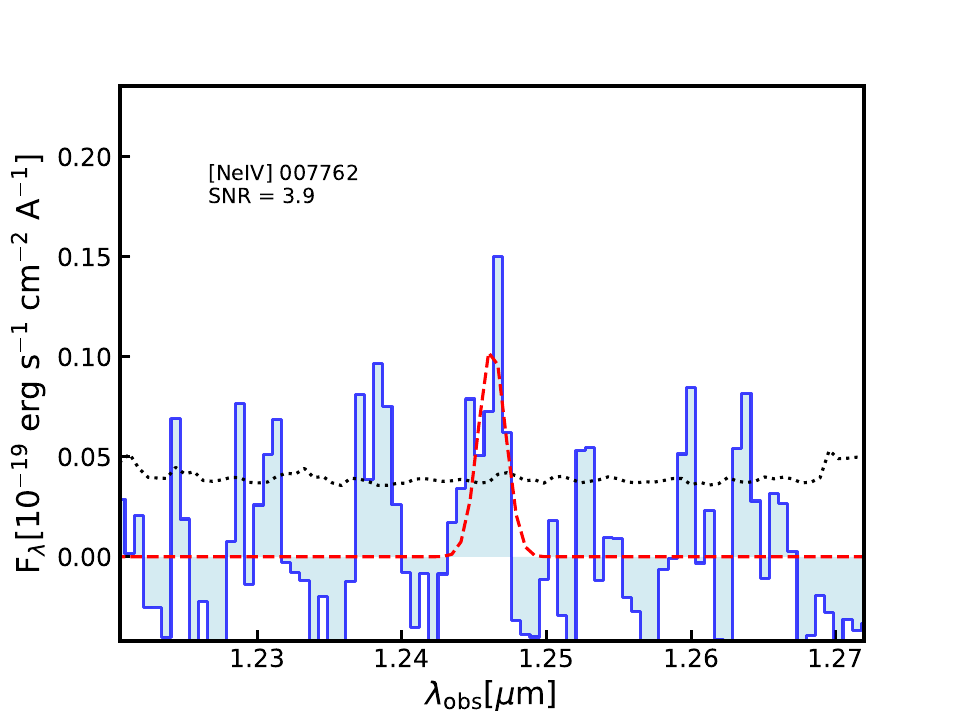}  
     \end{subfigure}
     \hfill
     \begin{subfigure}[b]{0.285\paperwidth}
         \centering
         \includegraphics[width=0.28\paperwidth]{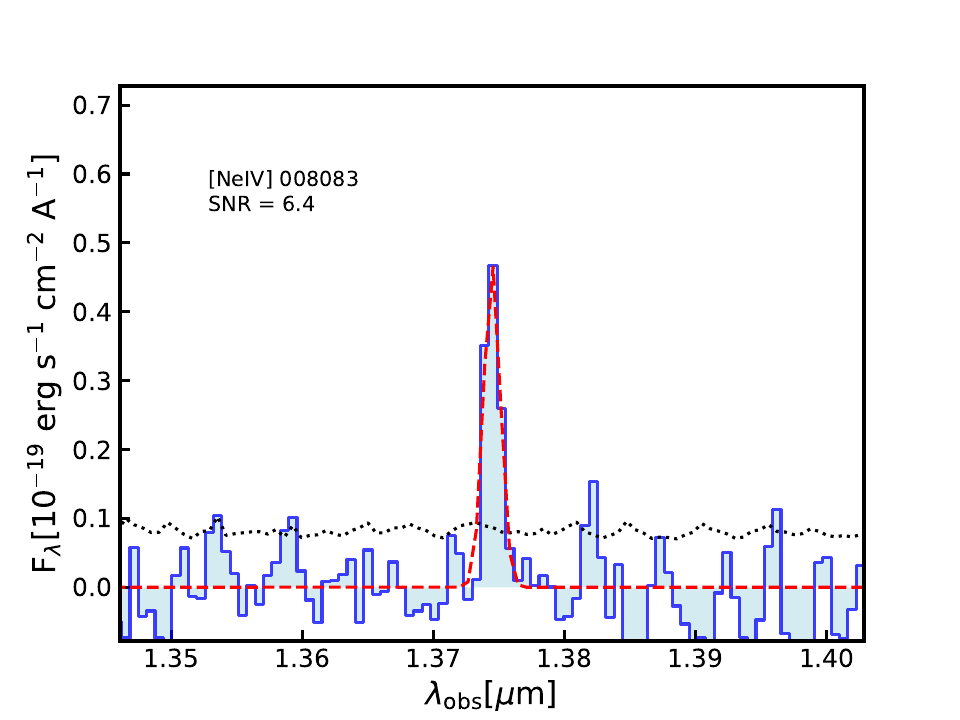}  
     \end{subfigure}
     \hfill
     \begin{subfigure}[b]{0.285\paperwidth}
         \centering
         \includegraphics[width=0.28\paperwidth]{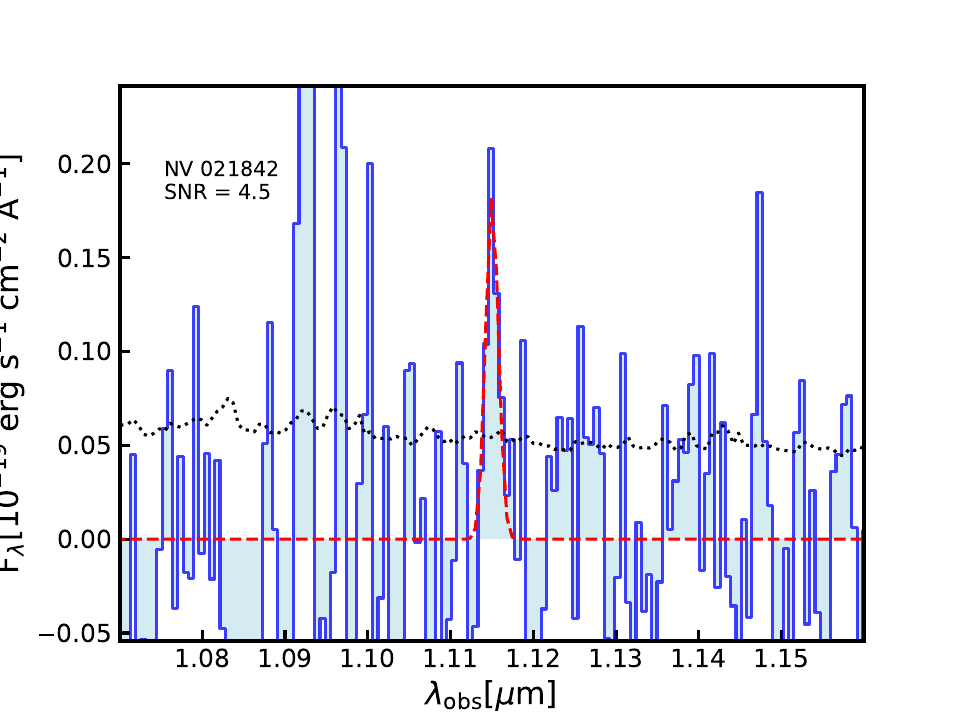}  
     \end{subfigure}

    \centering
     \begin{subfigure}[b]{0.285\paperwidth}
         \centering
         \includegraphics[width=0.28\paperwidth]{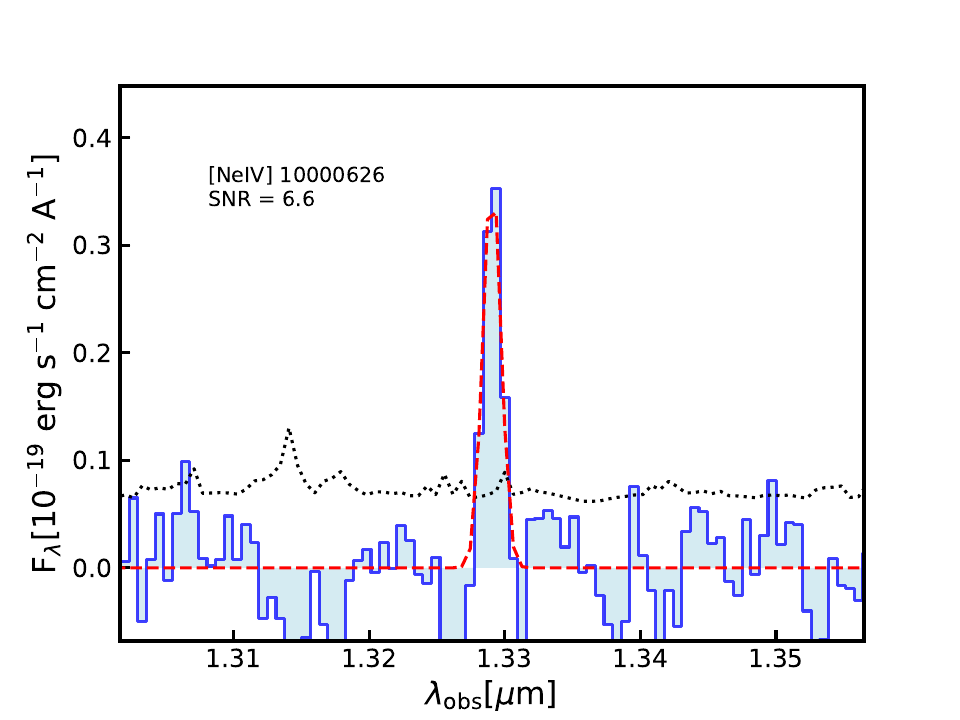}  
     \end{subfigure}
     \hfill
     \begin{subfigure}[b]{0.285\paperwidth}
         \centering
         \includegraphics[width=0.28\paperwidth]{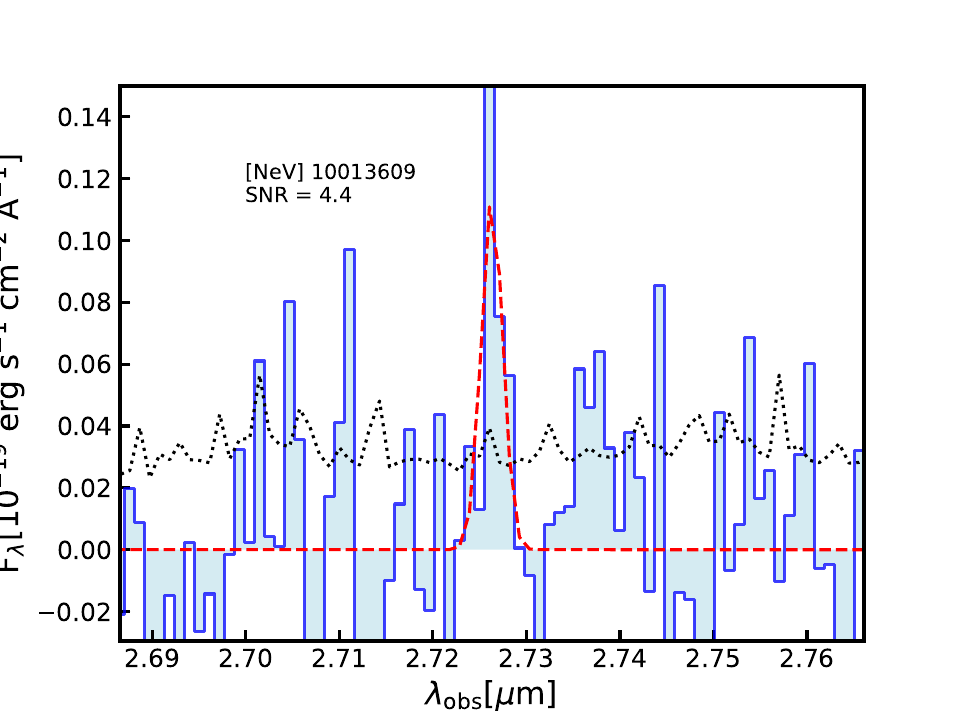}  
     \end{subfigure}
     \hfill
     \begin{subfigure}[b]{0.285\paperwidth}
         \centering
         \includegraphics[width=0.28\paperwidth]{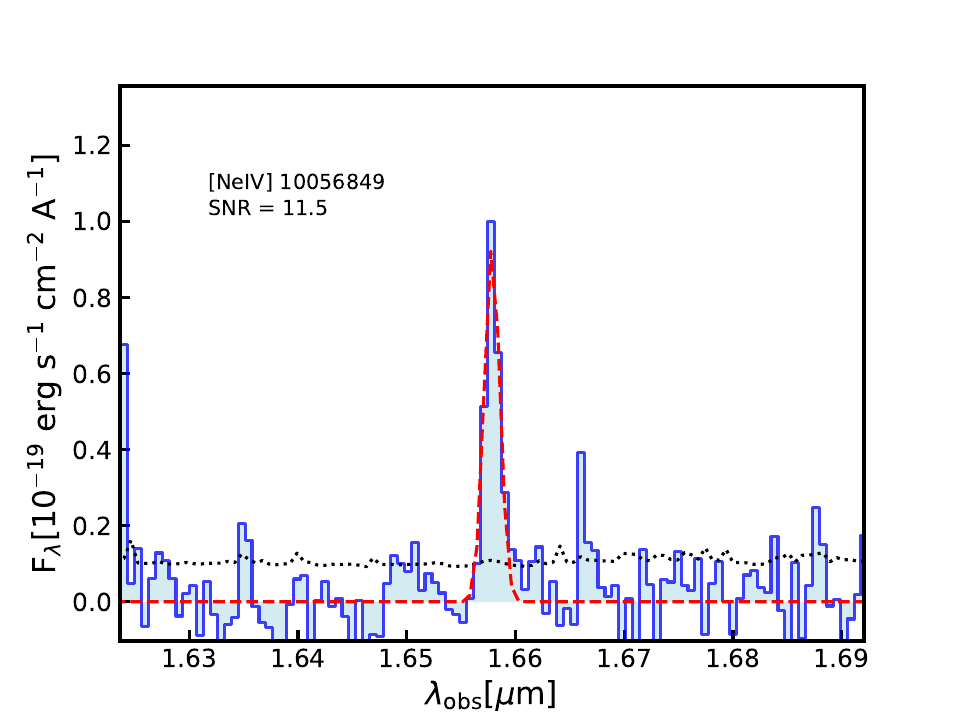}  
     \end{subfigure}
     \begin{subfigure}[b]{0.285\paperwidth}
         \includegraphics[width=0.28\paperwidth]{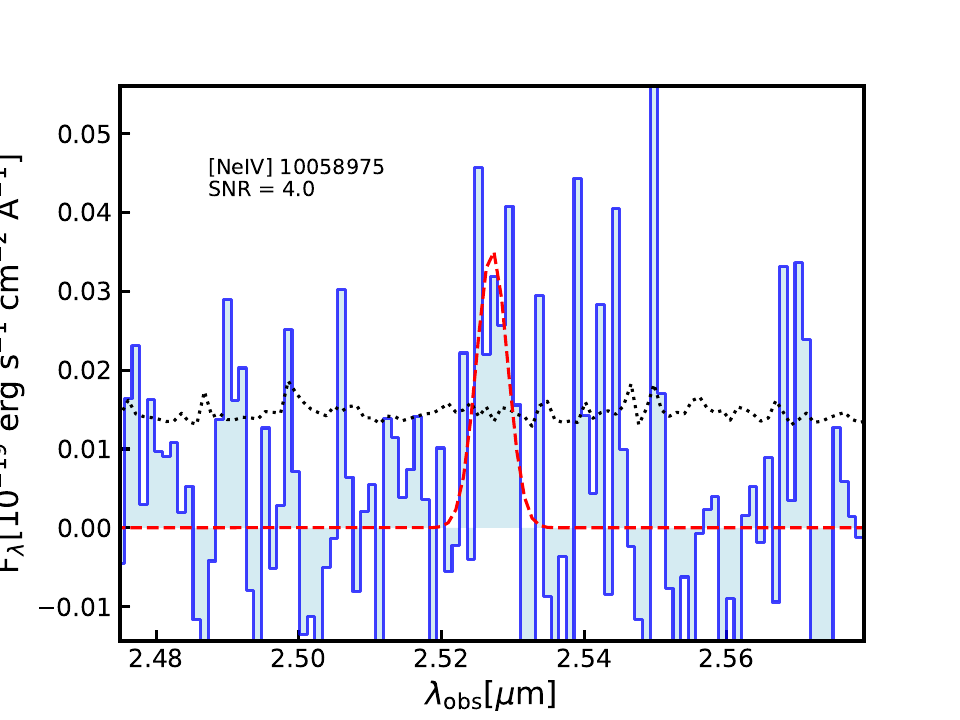}  
     \end{subfigure}
        \caption{Summary of the detection of \NeIVl2424, \NeVl3420 and NV$\lambda$1240. The blue lines show the continuum subtracted observed spectrum, black dotted lines indicate the uncertainties on the flux and the red dashed lines show the best fit to the data.  }
        \label{fig:High_ion}
\end{figure*}

We fit the \NeIVl2424, \NeVl3427 and NV$\lambda$1240 emission lines in the \S~\ref{sec:UV_fitting} and we show the spectra for the objects with detected \NeIVl2424 in Figure \ref{fig:High_ion}. Furthermore, we report the fluxes of the detected lines in Table \ref{table:Sample_high_ion}.

\begin{table*}
 \centering
   \caption{List of AGN and their detected fluxes of high ionisation lines: \NeIVl2422, \NeVl3420 and \NVl1240.}
    \begin{tabular}{@{}lcccc@{}} 
\hline 
\hline 
ID & field & Flux [NeIV]$\lambda$2424 & Flux [NeV]$\lambda$3427 & Flux NV$\lambda$1240\\
   &       & 10$^{-19}$ergs s$^{-1}$ cm$^{-2}$ & 10$^{-19}$ergs s$^{-1}$ cm$^{-2}$ & 10$^{-19}$ergs s$^{-1}$ cm$^{-2}$\\
 \\
\hline 
4902& 1210 & -& -& -\\
7099& 1210 & -& -& -\\
7762& 1210 & 2.38$\pm 0.67$& -& -\\
8083& 1210 & 7.50$\pm 1.17$& -& -\\
8456& 1210 & -& -& -\\
8880& 1210 & -& -& -\\
9422& 1210 & -& -& -\\
9452& 1210 & -& -& -\\
10073& 1210 & -& -& -\\
16745& 1210 & -& -& -\\
17072& 1210 & -& -& -\\
17670& 1210 & -& -& -\\
21842& 1210 & -& -& 3.48$\pm 0.17$\\
22251& 1210 & -& -& -\\
10000626& 1210 & 5.91$\pm 1.07$& -& -\\
10008071& 1210 & -& -& -\\
10011849& 1210 & -& -& -\\
10012477& 1210 & -& -& -\\
10012511& 1210 & -& -& -\\
10013597& 1210 & -& -& -\\
10013609& 1210 & -& 3.37$\pm 0.75$& -\\
10013905& 1210 & -& -& -\\
10015338& 1210 & -& -& -\\
10035295& 1210 & -& -& -\\
10036017& 1210 & -& -& -\\
10040620& 1210 & -& -& -\\
10056849& 1210 & 16.75$\pm 1.70$& -& -\\
10058975& 1210 & 1.88$\pm 0.48$& -& -\\
95256& 3215 & -& -& -\\
99671& 3215 & -& -& -\\
104075& 3215 & -& -& -\\
108487& 3215 & -& -& -\\
111091& 3215 & -& -& -\\
111511& 3215 & -& -& -\\
114573& 3215 & -& -& -\\
132213& 3215 & -& -& -\\
143403& 3215 & -& -& -\\
201127& 3215 & -& -& -\\
202208& 3215 & -& -& -\\
208643& 3215 & -& -& -\\
209979& 3215 & -& -& -\\
\hline 
\end{tabular} 

  \label{table:Sample_high_ion}
  \\
\end{table*}

\section{Emission lines used in this work}

In Table \ref{table:eml} we summarise the emission lines used in this work, their wavelength and ionisation potential. 

\begin{table}
 \centering
   \caption{List of emission lines used in this work, their wavelengths and ionisation potential}
 \begin{tabular}{lcc}
  \hline
  Emission line & $\lambda_{\rm rest}$(\AA) & Ionisation potential (eV)\\
  \hline
  \SII & 6718,32 & 10.4\\
  \NII & 6584,48 & 14.5\\
  \Ha & 6563 &  13.6\\
  \OIII & 5008,4961 & 35.1\\
  \Hb & 4861 & 13.6\\
  \HeII & 4686 & 54.4\\
  \OII & 3727,29 & 13.6\\
  \NeIII &  3689  & 41.0 \\
  \NeV & 3427 & 97.2\\
  \NeIV & 2422,24 & 63.4\\
  \CIII & 1907,09 & 24.4 \\
  \HeII & 1640 & 54.4\\
  \CIV & 1548,1550 & 47.9 \\
  \NVs & 1239,42 & 77.5 \\
 \hline
 \end{tabular}
  \label{table:eml}
\end{table}
\end{appendix}
\end{document}